\documentclass[12pt]{article}
\usepackage{epsf}
\usepackage{amsmath}
\usepackage{amsfonts}
\usepackage{amssymb}
\usepackage{graphicx}
\usepackage{color}
\usepackage{psfrag}
\usepackage{cite}
\usepackage{hyperref}

\usepackage{ifpdf}

\newcommand{\bmat}{\left(\begin{array}}
\newcommand{\emat}{\end{array}\right)}

\def\yzero{\smash{\hbox{$y\kern-4pt\raise1pt\hbox{${}^\circ$}$}}}

\def\beq{\begin{equation}}
\def\eeq{\end{equation}}
\def\beqa{\begin{eqnarray}}
\def\eeqa{\end{eqnarray}}

\def\-{\hphantom{-}}
\def\ov{\overline}
\def\s2{\frac{1}{\sqrt2}}

\def\beq{\begin{equation}}
\def\eeq{\end{equation}}
\def\beqa{\begin{eqnarray}}
\def\eeqa{\end{eqnarray}}

\def\diag{{\rm diag \,}}
\def\IF{\relax{\rm I\kern-.18em F}}
\def\II{\relax{\rm I\kern-.18em I}}

\def\Dsl{\,\raise.15ex\hbox{/}\mkern-13.5mu D} 

\def\IC{{\bf C}}
\def\IS{{\bf S}}
\def\IR{{\bf R}}
\def\IZ{{\bf Z}}
\def\IX{{\bf X}}
\def\IY{{\bf Y}}

\def\IT{{\bf T}}
\def\IP{\bf P}

\def\wDseven{{\widehat{\rm D7}}}
\def\wDzero{{\widehat{\rm D0}}}
\def\aDnine{{\overline{{\rm D}9}}}

\def\A{{\bf A}}

\newcommand{\drawsquare}[2]{\hbox{%
\rule{#2pt}{#1pt}\hskip-#2pt
\rule{#1pt}{#2pt}\hskip-#1pt
\rule[#1pt]{#1pt}{#2pt}}\rule[#1pt]{#2pt}{#2pt}\hskip-#2pt
\rule{#2pt}{#1pt}}

\newcommand{\fund}{\raisebox{-.5pt}{\drawsquare{6.5}{0.4}}}
\newcommand{\Ysymm}{\raisebox{-.5pt}{\drawsquare{6.5}{0.4}}\hskip-0.4pt%
        \raisebox{-.5pt}{\drawsquare{6.5}{0.4}}}
\newcommand{\Yasymm}{\raisebox{-3.5pt}{\drawsquare{6.5}{0.4}}\hskip-6.9pt%
        \raisebox{3pt}{\drawsquare{6.5}{0.4}}}

\catcode`\@=11   
\newdimen\@rotdimen
\newbox\@rotbox  

\def\@vspec#1{\special{ps:#1}}
\def\@rotstart#1{\@vspec{gsave currentpoint currentpoint translate
   #1 neg exch neg exch translate}}
\def\@rotfinish{\@vspec{currentpoint grestore moveto}}
\def\@rotr#1{\@rotdimen=\ht#1\advance\@rotdimen by\dp#1%
   \hbox to\@rotdimen{\hskip\ht#1\vbox to\wd#1{\@rotstart{90 rotate}%
   \box#1\vss}\hss}\@rotfinish}
\def\@rotl#1{\@rotdimen=\ht#1\advance\@rotdimen by\dp#1%
   \hbox to\@rotdimen{\vbox to\wd#1{\vskip\wd#1\@rotstart{270 rotate}%
   \box#1\vss}\hss}\@rotfinish}%
\def\@rotu#1{\@rotdimen=\ht#1\advance\@rotdimen by\dp#1%
   \hbox to\wd#1{\hskip\wd#1\vbox to\@rotdimen{\vskip\@rotdimen
   \@rotstart{-1 dup scale}\box#1\vss}\hss}\@rotfinish}%
\def\@rotf#1{\hbox to\wd#1{\hskip\wd#1\@rotstart{-1 1 scale}%
   \box#1\hss}\@rotfinish}%
\def\rotate{\@ifnextchar[{\@rotate}{\@rotate[l]}}
\def\@rotate[#1]#2{\setbox\@rotbox=\hbox{#2}\@nameuse{@rot#1}\@rotbox}

\catcode`\@=12

\topmargin
-1.5cm
\textwidth
15.5cm
\textheight
23.5cm
\oddsidemargin
0.7cm
\evensidemargin
1.2cm

\begin{document}

\makeatletter
\@addtoreset{equation}{section}
\makeatother
\renewcommand{\theequation}{\thesection.\arabic{equation}}
\renewcommand{\thefootnote}{\fnsymbol{footnote}}
\pagestyle{empty}
\rightline{ IFT-UAM/CSIC-13-063}
\vspace{0.1cm}
\begin{center}
\LARGE{\bf Discrete gauge symmetries from\\ (closed string) tachyon condensation \\[12mm]}
\large{M. Berasaluce-Gonz\'alez$^{1,2}$\footnote{mikel.berasaluce@csic.es}, M. Montero$^{1,2}$\footnote{mig.montero@estudiante.uam.es}, \\A. Retolaza$^{1,2}$\footnote{ander.retolaza@uam.es}, A. M. Uranga$^2$\footnote{angel.uranga@uam.es}\\[3mm]}
\footnotesize{${}^{1}$ Departamento de F\'{\i}sica Te\'orica, Facultad de Ciencias\\[-0.3em] 
Universidad Aut\'onoma de Madrid, 28049 Madrid\\
${}^2$ Instituto de F\'{\i}sica Te\'orica IFT-UAM/CSIC,\\[-0.3em] 
C/ Nicol\'as Cabrera 13-15, Universidad Aut\'onoma de Madrid, 28049 Madrid, Spain} \\ 

\vspace*{2.5cm}

\small{\bf Abstract} \\[5mm]
\end{center}
\begin{center}
\begin{minipage}[h]{16.0cm}
The study of discrete gauge symmetries in field theory and string theory is often carried out by embedding them into continuous symmetries. Many symmetries however do not seem to admit such embedding, for instance discrete isometries given by large diffeomorphisms in compactifications. We show that in the context of string theory even those symmetries can be embedded into continuous ones. This requires extending the system to a supercritical string theory configuration with extra dimensions, on which the continuous symmetry acts. The extra dimensions are subsequently removed by closed string tachyon condensation, which breaks the continuous symmetry but preserves a discrete subgroup. The construction is explicit and the tachyon condensation can even be followed quantitatively for lightlike tachyon profiles. The embedding of discrete into continuous symmetries allows a realization of charged topological defects as closed string tachyon solitons, in tantalizing reminiscence of the construction of D-branes as open tachyon solitons.

\end{minipage}
\end{center}
\newpage
\setcounter{page}{1}
\pagestyle{plain}
\renewcommand{\thefootnote}{\arabic{footnote}}
\setcounter{footnote}{0}

\tableofcontents

\vspace*{1cm}

\section{Introduction}


Discrete symmetries are a key ingredient in particle physics, and especially in physics beyond the Standard Model, for instance R-parity in supersymmetric extensions of the Standard Model or discrete symmetries in flavour physics. Beyond this phenomenological relevance, it is important to investigate the nature of discrete symmetries at a more fundamental level. Indeed, there are strong suggestions that global symmetries, either continuous or discrete, cannot exist in consistent quantum theories including gravity, such as string theory (see \cite{Banks:1988yz,Abbott:1989jw,Coleman:1989zu} for early viewpoints, and e.g.\cite{Kallosh:1995hi,Banks:2010zn} and references therein, for more recent discussions); hence, {\em exact} discrete symmetries should have a gauge nature \cite{Alford:1988sj,Krauss:1988zc,Preskill:1990bm}.

The realization of discrete gauge symmetries in string theory has been recently explored in several papers \cite{Camara:2011jg,BerasaluceGonzalez:2011wy, BerasaluceGonzalez:2012vb,BerasaluceGonzalez:2012zn}\footnote{See also \cite{Gukov:1998kn,Burrington:2006uu} and \cite{Ibanez:2012wg,Anastasopoulos:2012zu,Honecker:2013hda,fernando} for related applications, and \cite{Kobayashi:2006wq,Lebedev:2007hv,Nilles:2012cy} for discrete symmetries in heterotic orbifolds.}. In these papers,  the discrete symmetries arise as subgroups of continuous gauge symmetries broken by their gauging of (or Higgsing by) certain scalars of the theory. The prototypical case is that of $\IZ_n$ discrete gauge symmetries from $U(1)$ gauge groups on D-branes, broken by 4d St\"uckelberg couplings. This appearance of discrete symmetries from continuous ones is extremely useful, for instance to make contact with the 4d field theory description,  and to construct charged particles and other topological defects \cite{Banks:2010zn} (or, even in purely field theoretical setups, to study anomaly cancellation conditions  
\cite{Ibanez:1991hv,Ibanez:1991pr,Banks:1991xj,Ibanez:1992ji}). An important point, however, is that the scale of the massive gauge bosons is of the order of the string scale; from the 4d effective theory viewpoint, the continuous symmetry can be regarded as a useful device that facilitates the understanding of the discrete gauge symmetry, at the price of `integrating in' a sector of string scale physics, relevant to the topology of the symmetry breaking and the charged objects.

In string theory, either in 10d or in compactifications, there are however many examples of discrete symmetries that cannot be embedded into a continuous group. A prototypical example in compactifications is that of discrete gauge symmetries from `large' discrete isometries of the compactification space (i.e. corresponding to diffeomorphisms which cannot be continuously connected to the identity). The purpose of this paper is to realize that such discrete symmetries can actually be embedded into continuous ones, if one is ready to `integrate in' a suitable sector of stringy physics. We will argue that this suitable sector is provided by extra spacetime dimensions, on which the discrete symmetry can be promoted to a continuous one. As we will explain, extending the theories beyond the familiar 26d or 10d, for the bosonic or superstring theories, requires considering supercritical string theory models \cite{Chamseddine:1991qu}. 

The analogue of the gauging/Higgsing mechanism is condensation of the closed string tachyon in the spectrum of the supercritical theory. This removes the extra dimensions to recover the original critical theory, and breaks the continuous gauge symmetry down to the discrete group. Although the closed tachyon condensation involves string scale physics, this was also the case in the familiar setup of breaking continuous groups by gauging/Higgsing. In fact, as in the latter case, the effects of closed tachyon condensation with respect to the symmetry and its breaking are topological and can be analyzed reliably. One important implication is that discretely charged topological defects can be constructed as solitons of the closed string tachyon field, in a way reminiscent of the construction of D-branes as (open string) tachyon solitons, and its connection with K-theory. Note however that the dynamical details of closed tachyon condensation are very different from a gauging/Higgsing; hence, another take on this paper is the proposal of a new dynamical mechanism to realize discrete gauge symmetries from continuous ones, beyond those considered hitherto.

The paper is organized as follows. In the next subsection we warm up with an analogous realization of discrete symmetries, but in the more familiar open string setup: a $\IZ_2$ symmetry in type I theory, realized as a subgroup of a continuous $U(1)$ broken by open string tachyon condensation. In Section \ref{sec:supercritical} we review the construction of different supercritical string theories and their closed tachyon condensation decays to the familiar bosonic, heterotic and type II counterparts in the critical dimension  \cite{Hellerman:2004zm,Hellerman:2006ff}. In Section \ref{sec:rotations} we consider examples of discrete $\IZ_n$ symmetries of string models in critical dimension and their embedding as rotations in extra dimensions in supercritical extensions, e.g. spacetime parity in section \ref{sec:parity}, and a $\IZ_2$ symmetry of the $SO(32)$ 10d heterotic in section \ref{sec:hetz2}. In section \ref{sec:defects}, we describe the construction of $\IZ_n$ charged topological defects as closed tachyon solitons. In Section \ref{sec:translations}, we propose a more general embedding of discrete symmetries, in which the continuous group acts as a translation in a compact $\IS^1$ supercritical dimension, along which the theory picks up a $\IZ_n$ holonomy. The resulting configuration is introduced in section \ref{sec:mapping-torus}, and in section \ref{sec:superimposing} it is related to the description of discrete gauge symmetries as a sum over disconnected theories \cite{Hellerman:2010fv}. The closed tachyon solitons describing $\IZ_n$ defects are discussed in section \ref{sec:fluxbranes}. In section \ref{sec:examples} we consider this realizations of different examples of discrete symmetries, including large isometries. In section \ref{sec:dual} we consider further examples of non-geometric discrete symmetries, which actually turn into large isometries upon use of T-duality. In section \ref{sec:nonabelian} we discuss a generalization to non-abelian discrete gauge symmetries, and present an explicit example of discrete Heisenberg groups. Section \ref{sec:final} contains our conclusions. Appendix \ref{sec:partition-functions} contains the partition functions for the different supercritical strings, while appendix \ref{sec:tori-quintic} describes examples of large isometries for $\IT^2$ and CY threefolds. Appendix \ref{cosets} provides a formal description of the non-abelian case, and applies it to the example of section \ref{sec:nonabelian}.

\subsection{Warmup: a type I $\IZ_2$ symmetry from open string tachyon condensation}
\label{sec:typeiz2}

To clarify the proposal, in this section we present an analogous implementation in the context of \textit{open} string tachyon condensation. We use it in 10d type I theory to derive a discrete $\IZ_2$ gauge symmetry from a continuous one. Although the argument is simple, to our knowledge it has not appeared in the literature (see \cite{Collinucci:2008pf} for a similar phenomenon in type IIB orientifolds).

Recall that in 10d type I theory there exist several $\IZ_2$ charged non-BPS branes \cite{Witten:1998cd}. In particular we focus on the D7- and D0-branes (denoted by $\wDseven$- and $\wDzero$-branes in what follows), which pick up a $-1$ when moved around each other \cite{Witten:1998cd,Gukov:1999yn}. Hence, they correspond to a $\IZ_2$ charged particle and the dual  codimension-2 $\IZ_2$ charged defect (the 10d analogue of 4d $\IZ_2$ string), associated to a $\IZ_2$ discrete gauge symmetry. Since the $\wDzero$-brane is a spinor under the $SO(32)$ perturbative type I gauge symmetry \cite{Sen:1998tt,Sen:1998ki}, the $\IZ_2$ can be defined as acting as $-1$ on spinors and leaving tensors invariant. 

Naively, this type I $\IZ_2$ symmetry cannot be described as a discrete remnant of a broken continuous gauge symmetry. However, this can be achieved by regarding type I theory as the endpoint of open string tachyon condensation, starting from a configuration with additional D9-$\aDnine$ brane pairs \cite{Sugimoto:1999tx}; this is natural given the construction of non-BPS branes in terms of brane-antibrane pairs. Consider e.g. the case of two extra D9-$\aDnine$ brane pairs. The gauge symmetry is enhanced to $SO(34)\times SO(2)$, and there is a complex tachyon in the bifundamental, i.e. a vector of $SO(34)$ with $SO(2)\simeq U(1)$ charge $+1$ (and $-1$ for the conjugate scalar)\footnote{\label{fermioncontent}For future convenience, we mention that in models with $n$ extra D9-$\aDnine$ pairs, there are massless fermions transforming under $SO(32+n)\times SO(n)$ as follows \cite{Sugimoto:1999tx}: one set of chiral spinors in the representation $(\Yasymm,1)+(1,\Ysymm)$, and one opposite-chirality spinor in the $(\fund,\fund)$.}. On the other hand, the $\wDzero$-brane can be shown\footnote{This follows from the quantization of the fermion zero modes in $\wDzero$-D9 and $\wDzero$-$\aDnine$ sectors, and restricting onto states invariant under the world-volume $O(1)$ gauge symmetry.} to transform as a chiral bi-spinor, namely a chiral $SO(34)$ spinor with $U(1)$ charge $+ \frac 12$, and an opposite-chirality $SO(34)$ spinor with $U(1)$ charge $-\frac 12$. Tachyon condensation imposes the breaking 
\beqa
SO(34)\times SO(2) \rightarrow SO(32)\times SO(2)'\times SO(2)\stackrel{\langle T\rangle}{\longrightarrow} SO(32)\times SO(2)_{\rm diag}
\eeqa
(The intermediate step just displays the two $SO(2)$ symmetries most relevant in the final breaking). The phenomenon is very similar to a Higgs mechanism, with the {\em proviso} that the diagonal subgroup actually disappears from the theory
(this is analogous to the disappearance of the diagonal $U(1)$ in the annihilation of spacetime filling brane-antibrane pairs, see e.g. \cite{Bergman:2000xf} for discussions). The anti-diagonal combination $U(1)_{\rm anti}$ of $SO(2)'\times SO(2)$,  generated by $Q_{\rm anti}=Q_{SO(2)}-Q_{SO(2)'}$, is Higgsed down by the tachyon, which carries charge $+2$, thereby leaving a remnant $\IZ_2$ discrete gauge symmetry. The $\IZ_2$ charged particles are the $\wDzero$-brane states, which transform as a chiral  $SO(32)$ spinor with $U(1)_{\rm anti}$ charge $+1$. 

The main lesson is that this $\IZ_2$ symmetry of 10d type I theory\footnote{Incidentally, other $\IZ_2$ charged branes can be associated to discrete $\IZ_2$ subgroups of continuous symmetries, albeit associated not to gauge bosons but to higher RR $p$-forms, when described as K-theory valued objects \cite{Moore:1999gb}.} can be derived as an unbroken discrete symmetry of a continuous gauge symmetry, by regarding the theory as the endpoint of (open string) tachyon condensation. The `parent' theory includes extra degrees of freedom on which the continuous symmetry acts, and which disappear upon tachyon condensation. The purpose of this paper is to develop a similar description for other symmetries, which involves extra spacetime dimensions and closed string tachyon condensation. In fact, in section \ref{sec:hetz2} we will re-encounter the above type I pattern in the $SO(32)$ heterotic theory when regarded as the endpoint of {\em closed} string tachyon condensation (actually in agreement with a duality proposed in \cite{Hellerman:2004zm}).

\section{Supercritical strings and dimension quenching}
\label{sec:supercritical}

In the above type I example, the $\IZ_2$ symmetry acts on gauge quantum numbers, so its embedding in a continuous symmetry simply demands enlarging the gauge group. This is achieved with extra brane-antibrane pairs, removed by open string tachyons. This is natural since the $\IZ_2$ charged objects are D-branes, whose charges lie in K-theory. 

Clearly other discrete symmetries are radically different, and may involve non-trivial geometric actions. A prototypical example are discrete isometries given by large diffeomorphisms (i.e. not in the connected component of the  identity), dubbed `discrete large isometries' from now on. Their embedding in a continuous group demands enlarging the underlying spacetime symmetry. We must enlarge the number of dimensions beyond the familiar $D=26,10$ for bosonic or superstring theories, and consider physical processes removing them (but leaving an unbroken discrete symmetry). This motivates the use of supercritical string theories, and their closed string tachyon condensation processes, reviewed in this section. 

Supercritical strings are defined by generalizing the worldsheet field content of the familiar 26d bosonic string, or 10d superstring, to $D$ dimensions (with extra changes for supercritical heterotics), and introducing a linear dilaton background, $\phi=\phi_0+V_\mu X^\mu$, to maintain the correct matter central charge 
\beqa
c_{\rm matt.}=D+6\alpha' V_\mu V^\mu=26\quad ,\quad c_{\rm matt.}=\frac 32 D+6\alpha' V_\mu V^\mu=15
\eeqa
for the bosonic and superstring theories. We consider a timelike linear dilaton in order to produce supercritical (rather than subcritical) theories, i.e. $D>26,10$ dimensions, and choose coordinates such that $V^\mu=0$ for $\mu\neq 0$.

As usual, consistent theories must fulfill the requirement of modular invariance, which leads to different supercritical theories, discussed below. 
The pattern of GSO-like projections (when required) determines the spacetime spectrum, in particular massless\footnote{Since the dilaton background breaks Poincar\'e invariance, one should be careful in talking about mass. We follow the convention in \cite{Hellerman:2004zm} of meaning the mass term arising in the equations of motion of the spacetime field.} and tachyonic fields. The ubiquity of closed string tachyons will be an important aspect. 

Contrary to open string tachyons, closed string instabilities are less understood (see e.g. \cite{Adams:2001sv,Rabadan:2002zq,Uranga:2002ag,Hellerman:2004qa,Adams:2005rb,Green:2007tr} for some discussions). Fortunately, there is a quite precise description of the different effects of closed tachyon condensation in supercritical strings, which is even quantitative for lightlike tachyon profiles \cite{Hellerman:2006nx,Hellerman:2006ff,Hellerman:2006hf,Hellerman:2007fc}\footnote{For other works on light-like tachyon condensation, see \cite{Hellerman:2007zz,Hellerman:2007ym,Hellerman:2008wp}.}. In particular, they have been shown to trigger a reduction in the number of dimensions \cite{Hellerman:2006ff}, dubbed `dimension quenching' \cite{Hellerman:2007fc}. Moreover, the standard 26d bosonic string theory, the 10d $SO(32)$ heterotic, and the 10d type 0 and type II superstrings, can be regarded as the endpoint of closed string tachyon condensation of suitable supercritical string theories.

\subsection{Supercritical bosonic strings}
\label{sec:super-bosonic}

The supercritical bosonic string in $D$-dimensional spacetime is defined by $D$ worldsheet bosons $X^M$, $M=0,\ldots, D-1$, and an appropriate timelike linear dilaton. As in the familiar 26d theory, the light spectrum contains a (real) closed string tachyon $T(X)$, and massless graviton, 2-form and dilaton fields, $G_{MN}$, $B_{MN}$, $\phi$. 

Since the tachyon vertex operator is the identity, a tachyon background couples as a  worldsheet potential. Tachyon condensation can be  followed quantitatively for specific `light-like' profiles, in which the tachyon $T(X^+)$ depends on the direction $X^+$. It describes the dynamics close to the boundary of an expanding bubble interpolating between two vacua: the `parent' with no tachyon and the endpoint of tachyon condensation. We focus on tachyon profiles describing the disappearance of spacetime dimensions, see \cite{Hellerman:2006ff} for additional details. To remove e.g. one dimension $Y\equiv X^{D-1}$,  we deform the worldsheet action by a superposition of conformal operators
\beqa
T(X^+,Y)=\mu_0^2 \exp(\beta X^+) -\mu_k^2\cos(k Y)\exp(\beta_k X^+)
\label{before-limit}
\eeqa
The parameters $\beta$, $\beta_k$ are fixed to make the operators marginal (i.e. satisfy the tachyon spacetime equations of motion), and $\mu_0$, $\mu_k$ are tuned to achieve a stationary (yet not stable, in this bosonic case) endpoint.

The theory simplifies in the limit $k\to 0$ keeping fixed $\alpha' k^2\mu_k^2\equiv \mu^2$ and $\mu'\equiv \mu_0^2-\mu_k^2$. The tachyon profile essentially becomes
\beqa
T(X^+,Y)=\frac{\mu^2}{2\alpha'} \exp(\beta X^+) \, Y^{\, 2} 
\label{after-limit}
\eeqa
Here we have removed certain terms (denoted by $T_0$ in eq.(2.9) of \cite{Hellerman:2006ff}), since they describe a possible left-over tachyon background in the endpoint theory (which we tune to be zero), and contain a term canceling against an upcoming (worldsheet) quantum correction \cite{Hellerman:2006ff}.

At $X^+\to -\infty$ we have the parent bosonic theory in $D$ dimensions and vanishing tachyon profile, while at $X^+\to \infty$ the strings are pinned at $Y=0$, so spacetime effectively loses one dimension. Integrating out the worldsheet field $Y$ at late $X^+$ produces a quantum correction that readjusts the metric and dilaton background, correctly accounting for the change in the central charge. 

The bottom line is that the dimension $Y$ disappears via closed string tachyon condensation. It is straightforward to generalize to the disappearance of several dimensions, and in particular to decay down to  the familiar 26d bosonic string theory. 

The general lesson about closed string tachyon condensation is the reduction of spacetime dimensions onto the locus of vanishing tachyon. This lesson can be applied to more general tachyon profiles (see section \ref{sec:translations}) even if the corresponding worldsheet theory is not exactly solvable; in other words, the above quadratic profile (\ref{after-limit}) is a local approximation for any tachyon profile sufficiently near a simple zero, around which the background is otherwise trivial (except for the linear dilaton). From this perspective, the trigonometric profile (\ref{before-limit}) would lead to a periodic array of zeroes, and the purpose of the $k\to 0$ limit is to decouple them and extract an isolated zero.

\subsection{Supercritical heterotic strings}
\label{sec:super-heterotic}

Let us review the supercritical heterotic strings in $D=10+n$ dimensions, dubbed $HO^{+(n)}$ and $HO^{+(n) /}$ in \cite{Hellerman:2004zm}, to which we refer the reader for details. In both, the worldsheet theory generalizes the 10d $SO(32)$ heterotic in its fermionic formulation. There are $D$ (left- and right-moving) bosons $X^M$, $M=0,\ldots, D-1$, and  $D$ right-moving fermions ${\tilde \psi}^M$, all in the vector representation of the $SO(1,D-1)$ spacetime Lorentz group. In addition, there are $(32+n)$ left-moving fermions, which we split as $\lambda^a$, $a=1,\ldots, 32$ and $\chi^m$, $m=1,\ldots, n$ (the latter are required to cancel the 2d gravitational anomaly). As mentioned, a timelike linear dilaton is introduced to achieve the correct matter central charges. The two theories differ in their GSO projections (see appendix \ref{sec:partition-functions} for the partition functions), and their properties are discussed independently in the  following.

\begin{table}
\begin{center}
\begin{tabular}{ | c || c | c || c | c |}
  \hline
  \ &  \multicolumn{2}{|c||}{$HO^{+(n)}$} & $HO^{+(n) /}$ & $HO^{+(n) /}/\IZ_2$ \\
  \hline
    Field & $\quad \;\; g_1\quad\;\; $ & $\quad\;\; g_2\quad\;\;$ & $g_1 $ & $g_2$ \\ \hline
   $X^{\mu}$ & $+$ & $+$ & $+$  & $+$\\ \hline
   $X^{m}$ & $+$ & $+$ & $+$  & $-$\\ \hline
${\tilde \psi}^{\mu}$  & $+$ & $-$  & $-$ & $+$\\ \hline
${\tilde \psi}^{m}$  & $+$ & $-$  & $-$ & $-$\\ \hline
$ { \lambda }^{a} $  & $-$ & $+$ & $-$ & $-$ \\ \hline
${\chi}^{m}$  & $+$ & $-$  & $-$ & $-$ \\ 
\hline
\end{tabular}
\caption{Charge assignments for the GSO projections in the two supercritical heterotic theories. For convenience we use the indices $\mu=0,\ldots,  9$, and $m=9,\ldots D-1$. Note that the index $m$ for the $\chi$s labels their multiplicity, but it is not a Lorentz index.}
\label{table-gso}
\end{center}
\end{table}

\subsubsection*{$HO^{+(n)}$ theory}

In the $HO^{+(n)}$ theory there are two GSO-like projections, which can be described as orbifolds by the $\IZ_2$ actions $g_1$, $g_2$ in table \ref{table-gso}. We recall that in quantization on the cylinder, fermions odd under such an action $g$ are antiperiodic in the $g$-untwisted sector and periodic in the $g$-twisted sector. The GSO $g_1$ acts on the 32 left-moving fermions as in the 10d $SO(32)$ heterotic string, while $g_2$ is an extension of the standard GSO projection in the right-moving sector (as recovered by `forgetting' the ${\tilde \psi}^m$, $\chi^m$). The symmetry is $SO(1,9+n)_{X,{\tilde\psi}}\times SO(n)_{\chi}\times SO(32)_\lambda$. Tachyonic and massless states arise only in the $g_1$-untwisted sector, and are as follows 
\begin{center}
\begin{tabular}{ccccc}
\hline
Sector & State &   & $D$-dim. field & Comment\\
\hline
Unt. & $\chi_{-\frac 12}^m\, |0\rangle $ & $\rightarrow$ &  $T^m$ & Tachyons \nonumber\\
 &$\alpha_{-1}^M \, {\tilde \psi}^N_{-\frac 12}\, |0\rangle$ & $\quad \rightarrow \quad$ & $G_{MN}, B_{MN}, \phi$ & graviton, 2-form, dilaton\nonumber\\
 & $\lambda_{-\frac 12}^a\lambda_{-\frac 12}^b\, {\tilde \psi}^M_{-\frac 12}\, |0\rangle$ & $\rightarrow$ & $A_M^{ab}$ & $SO(32)$ gauge bosons\\
\hline
$g_2$-twist. & $\alpha_{-1}^M \, |{\rm spinor}\rangle$ & $\quad \rightarrow \quad$ & $\Psi^M_\alpha$ & {\rm Rarita-Schwinger + Dirac}\nonumber\\
 & $\lambda_{-\frac 12}^a\lambda_{-\frac 12}^b|{\rm spinor}\rangle$ & $\rightarrow$ & $\lambda_\alpha^{ab}$ & Spinor in adj. of $SO(32)$ \\
\hline
\end{tabular}
\end{center}
The tachyons $T^m$ are singlets under the $SO(32)$ gauge group. The spinor groundstates arise from fermion zero modes for ${\tilde \psi}^M$, $\chi^m$, and have fixed overall chirality under $SO(1,9+n)\times SO(n)$.

This theory is related to the 10d $SO(32)$ heterotic theory by (light-like) tachyon condensation quenching the $n$ extra dimensions. In this case, the tachyon background couples as a  worldsheet superpotential
\beqa
\Delta {\cal L}_{2d}\, =\,-\frac{1}{2\pi} \int d\theta^+\, \sum_m\Lambda^m\, T^m(X)\, 
\label{het-supo}
\eeqa
where $T^m$ are functions of the $(0,1)$ superfields $X^M+i\theta^+{\tilde\psi}^M$, and we have Fermi superfields $\Lambda^m=\chi^m+\theta^+ F^m$, with $F^m$ being auxiliary fields (see e.g. \cite{Zumino:1997hf}). Using the kinetic term to integrate out the latter, the extra terms in components are a worldsheet potential for the worldsheet bosons and Yukawa couplings for worldsheet fermions
\beqa
{\cal L}_{\rm pot}\sim 
\sum_m \,(\, T^m(X)\, )^2\, \quad ,\quad \,
 {\cal L}_{\rm Yuk}\sim
 \,\partial_M T^m(X)\,\chi^m\,{\tilde \psi^M}
\label{het-supo-comp-1}
\eeqa
To describe the disappearance of all dimensions $X^m$,  we consider a profile for the tachyons, sketchily given by
\beqa
T^{m}(X^+,X)=\mu_k^2\sin(k X^m)\exp(\beta_k X^+)\quad \rightarrow \quad T^{m}(X^+,X)=\mu^2 X^m\,\exp(\beta_k X^+)\nonumber\\
\label{tachyon-het-1}
\eeqa
The LHS describes the deformation by exponential operators (taken marginal by tuning $\beta_k$), while the RHS describes the configuration after a $k\to 0$ limit, similar to the earlier one in section \ref{sec:super-bosonic}.  At $X^+\to -\infty$ we have the theory in $D$ dimensions and vanishing tachyon profile, while at $X^+\to \infty$ the dynamics truncates to the slice $X^m=0$, since all the extra worldsheet fields are made massive by (\ref{het-supo-comp-1}). The endpoint of tachyon condensation is the 10d supersymmetric $SO(32)$ heterotic string theory\footnote{Actually, a BPS state in this 10d theory, c.f. \cite{Hellerman:2006ff}}. It is worth noting that from the spacetime perspective, the 10d fermions arise as zero modes of the $(10+n)$d fermions coupled to the tachyon kink.

We conclude by mentioning that, exactly as in the 10d heterotic theories, it is straightforward to modify the GSO projections of the 32 worldsheet fermions $\lambda^a$ to construct a supercritical version of the $E_8\times E_8$ heterotic theory.

\subsubsection*{$HO^{+(n)/}$ theory}

In the $HO^{+(n)/}$ theory, the GSO projection is given by the action $g_1$ in the third column of table \ref{table-gso}. All left-moving fermions $\lambda$, $\chi$ are on equal footing, so they are collectively denoted $\lambda^a$, $a=1,\ldots, 32+n$. The symmetry is $SO(1,9+n)_{X,{\tilde\psi}}\times SO(32+n)_\lambda$. Light states arise only in the $g_1$-untwisted sector, and are as follows
\begin{center}
\begin{tabular}{ccccc}
\hline
Sector & State &   & $D$-dim. field & Comment\\
\hline
$g_1$-untwisted & $\lambda_{-\frac 12}^a\, |0\rangle $ & $\rightarrow$ &  $T^a$ & Tachyons \nonumber\\
& $\alpha_{-1}^M \, {\tilde \psi}^N_{-\frac 12}\, |0\rangle$ & $\quad \rightarrow \quad$ & $G_{MN}, B_{MN}, \phi$ & graviton, 2-form, dilaton\nonumber\\
& $\lambda_{-\frac 12}^a\lambda_{-1/2}^b\, {\tilde \psi}^N_{-\frac 12}\, |0\rangle$ & $\rightarrow$ & $A_M^{ab}$ & $SO(32+n)$ gauge bosons\\
\hline
\end{tabular}
\end{center}
The tachyons $T^a$ transform in the vector representation of the $SO(32+n)$ gauge group.

\medskip

The above theory has no spacetime fermions, so it is not a supercritical extension of the supersymmetric $SO(32)$ heterotic. However, the latter is related to a $\IZ_2$ orbifold of the $HO^{+(n) /}$ theory, defined by the element $g_2$ in table \ref{table-gso}. Since this breaks the Lorentz symmetry down to $SO(1,9)\times SO(n)$, we use indices $\mu=0,\ldots, 9$, $m=10,\ldots, D-1$. 

The original $HO^{+(n) /}$ spectrum is the $g_2$-untwisted sector, so it propagates in $D$ dimensions, and must be projected onto $g_2$-invariant states. In particular, the tachyons $T^a$, as well as the `mixed tensors' $G_{m\mu}, B_{m\mu}$, are forced to vanish at the fixed locus $X^m=0$. The only additional massless states arise from the $g_1g_2$-twisted sector, and correspond to the following 10d massless fields localized at the fixed locus $X^m=0$
\begin{center}
\begin{tabular}{cccc}
\hline
State &   & 10d field & Comment\\
\hline
$\alpha_{-1}^\mu |{\rm spinor} \rangle$ & $\rightarrow$ & $\psi^\mu_\alpha$  & Gravitino+dilatino\nonumber \\
$\lambda_{-\frac 12}^a\lambda_{-\frac 12}^b|{\rm spinor}\rangle$ & $\rightarrow$ &$\chi_\alpha^{ab}$ & chiral fermion in $(\Yasymm, 1)$ \nonumber \\
$\lambda_{-\frac 12}^a \alpha_{-\frac 12}^m|{\rm spinor}'\rangle$ & $\rightarrow$ &$\chi_{\dot{\alpha}}$ & opp. ch. fermion in $(\fund,\fund)$\nonumber \\
$\alpha_{-\frac 12}^m\alpha_{-\frac 12}^n|{\rm spinor}\rangle$ & $\rightarrow$ &$\chi_\alpha^{mn}$ & chiral fermion in $(1,\Ysymm)$\\
\hline
\end{tabular}
\end{center}
The spinor groundstate arises from the fermion zero modes of ${\tilde \psi}^\mu$, and the 10d chirality is fixed by the GSO projection. The representations of the fermions is under the $SO(32+n)$ gauge group and the $SO(n)_{\rm rot}$ rotational group in the coordinates $X^m$. Regarding the latter as some kind of gauge group, the fermion content motivated ref. \cite{Hellerman:2004zm} to propose a duality with type I with $n$ D9-$\aDnine$ brane pairs.

\medskip

This orbifold configuration relates to the 10d $SO(32)$ heterotic theory by (light-like) tachyon condensation. The tachyon background couples as a  worldsheet superpotential
\beqa
\Delta {\cal L}_{2d}\, =\,-\frac{1}{2\pi} \int d\theta^+\, \sum_a\Lambda^a\, T^a(X)\, 
\label{het-supo-second}
\eeqa
(with Fermi superfields $\Lambda^a=\lambda^a+\theta^+ F^a$), and the analogs of (\ref{het-supo-comp-1}) provide the 2d scalar potential and fermion couplings. The removal of the dimensions $X^m$, $m=10,\ldots, D-1$, requires a profile for the tachyons $T^{a=23+m}$, i.e.  $a=33,\dots, 32+n$, sketchily
\beqa
T^{23+m}(X^+,X)=\mu_k^2\sin(k X^m)\exp(\beta_k X^+)\quad \rightarrow \quad T^{23+m}(X^+,X)=\mu^2 X^m\,\exp(\beta_k X^+)\nonumber\\
\label{tachyon-het-2}
\eeqa
before and after the familiar $k\to 0$ limit. Note that we must restrict to tachyon profiles invariant under the orbifold $\IZ_2$ symmetry $X^m\to -X^m$, $T^a\to -T^a$.

At $X^+\to -\infty$ we have the parent orbifold configuration with $D$ dimensions and vanishing tachyon, while at $X^+\to \infty$ the dynamics truncates to the slice $X^m=0$. The endpoint of tachyon condensation is the 10d supersymmetric $SO(32)$ heterotic theory\footnote{Actually, a BPS state in this 10d theory, c.f. \cite{Hellerman:2006ff}}.

\subsection{Supercritical type 0 strings and decay to type II}
\label{sec:super-type0}

In this Section we describe 10d type II theories as the endpoint of closed string tachyon condensation in some supercritical theory. Supercritical type II theories exist, but only in dimensions $D=8p+2$ (because of the specific structure of their modular invariant partition function). Fortunately, a more generic extension can be obtained by considering supercritical type 0 theories in $D=10+2p$ dimensions, and performing a suitable orbifold to relate them to the 10d type II theories  \cite{Hellerman:2006ff}, as we now review.

The supercritical type 0 theories in $D$ dimensions are described by $D$ worldsheet bosons $X^M$, and left and right fermions $\psi^M$, ${\tilde \psi}^M$, with $M=0,\ldots, D-1$, and the usual timelike linear dilaton background. There is a GSO projection, associated to  $(-1)^{F_w}$, where $F_w$ is total worldsheet fermion number, with two choices that correspond to the type 0A or 0B theories, as in the critical type 0 theories.

The spacetime spectrum contains a real  tachyon $T(X)$ given by the groundstate in the NSNS sector. The massless NSNS sector contains the $D$-dimensional graviton, 2-form and dilaton. In the RR sector, the states are the tensor products of left and right (non-chiral) spinor groundstates, decomposing as a bunch of $p$-forms (different in the 0A and 0B theories).

The tachyon couples as a worldsheet superpotential, 
\beqa
\Delta {\cal L}\, =\, \frac{i}{2\pi}\int d\theta^+d\theta^-{ T}(X)\, 
\label{supo-type0}
\eeqa
where $T(X)$ depends on the $(1,1)$ superfields $X^M+i\theta^-\psi^M+i\theta^+{\tilde \psi}^M+i\theta^-\theta^+ F^M$ (see e.g. \cite{Zumino:1997hf}). Upon integrating out the auxiliary fields $F^M$, the terms in components describe a worldsheet potential and Yukawa couplings
\beqa 
{\cal L}_{\rm pot}\sim \partial^M T(X)\partial_M T(X) \quad ,\quad {\cal L}_{\rm Yuk}\sim \partial_M\partial_N{\cal T}(X)\, {\tilde \psi}^M \psi^N
\label{wsupo}
\eeqa
Condensation of this tachyon can produce dimension quenching in type 0 theories, but cannot connect down to 10d type II theories. In order to achieve the latter, we must instead consider a $\IZ_2$ quotient of the above supercritical configuration.

In particular, we focus on even dimensions $D=10+2p$, and split the extra $2p$ coordinates in two sets, denoted by $X^m$, $X'{}^m$. For such even $D$, there is a global symmetry on the worldsheet, corresponding to left-moving worldsheet fermion number $(-1)^{F_{L_w}}$ (i.e. under which $\psi^M$ are odd and ${\tilde \psi}^M$ are even).  In the critical $D=10$ case, orbifolding by $(-1)^{F_{L_w}}$ produces the 10d type II theories (since this projection combines with the type 0 GSO to produce independent left and right GSO projections). In the supercritical case, modular invariance requires to mod out by $g\equiv (-1)^{F_{L_w}}\cdot {\cal R}$, where ${\cal R}$ is the spacetime $\IZ_2$ action $X'{}^m\to -X'{}^m$, $X^m\to X^m$. The $g$-untwisted sector corresponds to the ($\IZ_2$ projected) old NSNS and RR sectors, and describe fields in $D=10+2p$ dimensions, while the twisted sectors are NS-R and R-NS sectors localized at $X'{}^m=0$, i.e. in $10+p$ dimensions. The NSNS and RR sectors contain bosonic fields, whereas the NS-R and R-NS sectors contain fermion fields. The choices of 0A or 0B as starting point determine the IIA or IIB - like projections in this twisted sector.

Focusing on the latter, the massless fields correspond to $(10+p)$-dimensional  vector-spinor fields $\psi^M_\alpha$, where $M$ runs through $D=10+2p$ coordinates, and $\alpha$ denotes a bi-spinor of $SO(1,9+p)\times SO(p)$, with an overall chirality projection (i.e. the decomposition of an $SO(1,9+2p)$ chiral spinor). The vector-spinor field splits as a gravitino and a Weyl spinor, as usual.

These supercritical non-compact orbifolds can be connected with 10d type II theories by closed string tachyon condensation. The local worldsheet coupling to the tachyon background is still given by (\ref{supo-type0}) and (\ref{wsupo}). The superpotential must be $\IZ_2$-odd, so the tachyon is a superposition of marginal operators with sine dependence on $X'{}^m$'s. For instance, choosing a sine dependence also on $X^m$, and taking two extra dimensions for simplicity, we have
\beqa
{T}=\mu \exp(\beta X^+)\, \mu_{k,k'}\, \sin (k X)\, \sin (k' X')
\label{typeii-tach}
\eeqa
In the familiar $k\to 0$ limit, we have the tachyon profile $T\sim XX'$, or in general (setting some constants equal for simplicity)
\beqa
{T}=\mu \exp(\beta X^+) \sum_{m=1}^p\, X^mX'{}^m
\label{typeii-tach-limit}
\eeqa
This removes the coordinates $X^m$, $X'{}^m$ and produces a 10d superstring theory at $X^m=X'{}^m=0$. It contains NSNS, NSR, RNS and RR sectors with a type II GSO projection, i.e. we recover the 10d type II theories. In particular, notice that the tachyon vanishes at $X=X'=0$ and does not give any dynamical mode after condensation.

\subsection{Dimensional reduction vs dimension quenching}
\label{sec:rec-quench} 

We conclude the discussion several conceptual remarks: Dimension quenching is drastically different from Kaluza-Klein dimensional reduction. From the spacetime viewpoint, dimension quenching causes the extra dimensions to completely disappear from the theory. In particular, there remain no towers of extra-dimensional momentum modes. Even at the level of massless modes, its effect on spacetime bosons differs from a truncation to the zero mode sector; for instance, mixed components $G_{\mu m}$ disappear completely (whereas they can survive in KK dimensional reduction). However, it is important to emphasize that, dimension quenching {\em does} behave like dimensional reduction for massless spacetime fermions; indeed, the 10d spinors arise from zero modes of the higher-dimensional Dirac operator coupled to the tachyon background \cite{Hellerman:2004zm}. 

Hence, from the spacetime perspective, the natural order parameter measuring the tachyon background is the derived quantity coupling to the spacetime fermions, namely $\partial_m T^a$ in heterotic, c.f. (\ref{het-supo-comp-1}), and $\partial_{m}\partial_n T$ in type 0/II, c.f. (\ref{wsupo}). 
The symmetry breaking pattern is mostly encoded in the quantum numbers of this quantity. For instance, in section \ref{sec:super-heterotic} the background $\partial_m T^a$ breaks the $SO(32+n)\times SO(n)$ gauge and rotational symmetries down to $SO(32)$ (times a diagonal $SO(n)_{\rm diag}$). Similarly, in section \ref{sec:super-type0} the background $\partial_m\partial_n T$ breaks the $SO(n)\times SO(n)$ rotational symmetries (seemingly with a left-over diagonal $SO(n)_{\rm diag}$). Although the diagonal symmetries $SO(n)_{\rm diag}$ would seem unbroken from a Higgsing perspective in spacetime, the microscopic worldsheet computation shows that they actually disappear. 

Let us end by remarking the close analogy of the last two paragraphs with similar discussions in the more familiar case of open string tachyon condensation (and their  differences/analogies with the Higgs mechanism c.f. section \ref{sec:typeiz2}). 

\section{Discrete gauge symmetries as quenched rotations}
\label{sec:rotations}

In general, discrete symmetries cannot be regarded as a discrete subgroup of a continuous group action on the theory. This happens for instance for
discrete large isometries in compactifications, namely discrete isometries associated to {\em large} diffeomorphisms of the geometry. In compactifications, discrete isometries of the internal space become discrete gauge symmetries of the lower-dimensional theory. Hence, discrete gauge symmetries from large isometries cannot in principle be regarded as unbroken remnants of some continuous gauge symmetry.

In this section we actually show that  even such discrete symmetries can be embedded into continuous groups, which however act on extra dimensions in a supercritical extension of the theory. The continuous orbits come out of the critical spacetime slice, and are quenched by a closed string tachyon condensation process, which thus breaks the continuous symmetry, yet preserves the discrete one. As mentioned at the end of the previous section, this process is similar to a Higgs mechanism, but with some important differences. The appearance of extra dimensions is very intuitive in extending discrete isometries to a continuous action, since these symmetries are related to properties of the metric; still our constructions apply to fairly general discrete symmetries.

\subsection{Spacetime parity}
\label{sec:parity}

A prototypical example of large diffeomorphisms it that of orientation reversing actions, for instance, and most prominently, spacetime parity. Consider the action $(x^0,x^1,\ldots,x^{2n-1})\to (x^0,-x^1,\ldots,-x^{2n-1})$ in $2n$-dimensional Minkowski spacetime. It has determinant $-1$, so it lies in a disconnected component of the Lorentz group. For convenience, we equivalently focus on the action $x^{2n-1}\to -x^{2n-1}$ (with other coordinates invariant), which lies in the same component disconnected from the identity.

It is easy to embed this symmetry into a continuous one, by adding one extra dimension $x^{2n}$, and considering the $SO(2)$ rotation in the 2-plane $(x^{2n-1},x^{2n})$. This is also easily implemented in string theory, as we now show using the supercritical bosonic string (since superstrings are actually not parity-invariant, due to chiral fermions or Chern-Simons couplings; still, parity can be combined with other actions to yield symmetries of such theories, see section \ref{sec:cp}).

Consider the supercritical bosonic string theory with one extra dimension, denoted by $x^{26}$, with the appropriate timelike linear dilaton. The theory is invariant under continuous $SO(2)$ rotations in the 2-plane $(x^{25},x^{26})$. As in section \ref{sec:super-bosonic} we can connect it with the critical 26d bosonic theory by a closed string tachyon profile (\ref{after-limit})
\beqa
T(X^+,X^{26})=\frac{\mu^2}{2\alpha'} \exp(\beta X^+) \, (X^{26})^{ 2} 
\eeqa

At $X^+\to -\infty$ the tachyon vanishes and we have a 27d theory with a continuous $SO(2)$ rotational invariance in the 2-plane $(x^{25},x^{26})$. At $X^+\to \infty$, the onset of the tachyon truncates the dynamics to the slice $X^{26}=0$, breaking the $SO(2)$ symmetry to the $\IZ_2$ subgroup $X^{25}\to -X^{25}$. Hence the $\IZ_2$ parity symmetry can be regarded as a discrete subgroup of a continuous higher dimensional rotation group, broken by the tachyon condensation removing the extra dimension.

The analogy of this breaking with a Higgs mechanism can be emphasized by using polar coordinates, $W=X^{25}+iX^{26}=|W|e^{i\theta}$. Then $X^{26}\sim W-{\ov W}$ and 
\beqa
T\sim (X^{26})^2\sim W^2-2W{\ov W}+{\ov W}^2\sim e^{2i\theta}-2+e^{-2i\theta}
\label{tachyon-polar}
\eeqa
The tachyon background implies vevs only for modes of even charge under the $U(1)$ symmetry, hence a $\IZ_2$ symmetry remains. This description will be useful for the construction of $\IZ_2$ charged defects in section \ref{sec:defects}.

Finally, note that although this construction embeds the discrete group into a continuous one, there is no actual 26d $SO(2)$ gauge boson. This will be achieved in a slightly different construction in section \ref{sec:translations}. 

\subsection{A heterotic $\IZ_2$ from closed tachyon condensation}
\label{sec:hetz2}

In the 10d $SO(32)$ heterotic string, the gauge group is actually $Spin(32)/\IZ_2$, and there is a $\IZ_2$ symmetry under which the (massive) spinor states are odd, while fields in the adjoint are even\footnote{Recall that the $\IZ_2$ by which $Spin(32)$ is quotiented prevents the presence of states in the vector representation; also notice that this is {\em not} the $\IZ_2$ discrete gauge symmetry we are interested in.}. We now propose a realization of this $\IZ_2$ symmetry as a discrete remnant of a continuous $U(1)$ symmetry, exploiting the supercritical heterotic strings introduced in section \ref{sec:super-heterotic}. 

For that purpose, it suffices to focus on the case of  $D=12$, i.e. two extra dimensions, denoted $x^{10}$, $x^{11}$. We complexify the extra worldsheet fields into a complex scalar $Z\equiv X^{10}+iX^{11}$, and a complex fermion $\Lambda=\lambda^{33}+i\lambda^{34}$. We consider the $U(1)$ action \beqa
Z \to e^{i\alpha} Z\quad  ,\quad \Lambda\to e^{-i\alpha}\Lambda
\eeqa
namely, the anti-diagonal $SO(2)_{\rm anti}\subset SO(2)_{\rm rot}\times SO(2)_{\rm gauge}\subset SO(2)_{\rm rot}\times SO(2+32)$. Consider now the tachyon background (\ref{tachyon-het-2}) producing the $SO(32)$ heterotic
\beqa
T^{33}(X)\sim e^{\beta X^+}\, X^{10}\quad , \quad T^{34}(X)\sim e^{\beta X^+}\, X^{11} \ \ \rightarrow \ \ T(X) \sim e^{\beta X^+} Z
\eeqa
which we have recast in terms of a holomorphic complex tachyon $T\equiv T^{33}+iT^{34}$. The order parameter $\partial_Z T$ transforms in the bifundamental of $SO(2)_{\rm rot}\times SO(2)_{\rm gauge}$, and breaks the $SO(2)_{\rm rot}\times SO(34)$ symmetry down to $SO(32)$ (times a diagonal factor which `disappears'). The anti-diagonal $SO(2)_{\rm anti}$, generated by $Q_{\rm anti}=Q_{SO(2)_{\rm gauge}}-Q_{SO(2)_{\rm rot}}$, is broken by a charge $+ 2$ the tachyon background. To show that there is an unbroken $\IZ_2$ acting as $-1$ on the $SO(32)$ spinors, it suffices to show that they descend from states with $SO(2)_{\rm anti}$ charge $\pm 1$. Indeed, they descend from the massive groundstates in the $g_2$ twisted (and $g_1$-untwisted) sector, which has $\lambda^a$, $\psi^m$ fermion zero modes; the states transform as a $SO(34)\times SO(2)_{\rm rot}$ bi-spinor, hence have $SO(2)_{\rm anti}$ charge $\pm 1$, and descend to $\IZ_2$ odd $SO(32)$ spinors. Note that the discussion parallels that of the type I $\IZ_2$ in section \ref{sec:typeiz2} (as suggested by the duality proposed in \cite{Hellerman:2004zm}).

A slightly unsatisfactory aspect of the construction is that there are actually no gauge bosons associated to the $SO(n)_{\rm rot}$ group. This could be achieved by curving the geometry of the extra dimensions in the radial direction, with the angular coordinates asymptoting to a finite size $\IS^{n-1}$.
Its $SO(n)$ isometry group would then produce an $SO(n)$ gauge symmetry (in 11d). Although the geometric curvature will render the worldsheet theory non-solvable, we expect basic intuitions of the flat space case to extend to the curved situation, in what concerns the relevant topology of symmetry breaking. In any event, we will eventually turn to a more general construction, with physical gauge bosons, in section \ref{sec:translations}.

\subsection{Topological defects from closed tachyon condensation}
\label{sec:defects}

In the embedding of a discrete symmetry into a continuous one acting on extra dimensions, all relevant degrees of freedom are eventually removed by the closed string tachyon condensation. We may therefore ask what we gain by such construction. The answer is that, in analogy with open string tachyon condensation, certain branes of the final theory can be constructed as solitons of the tachyon field. Since our focus is on aspects related to $\IZ_n$ discrete gauge symmetries, in this section we describe the tachyon profiles corresponding to the codimension-2 $\IZ_n$ charged defects (real codimension-2 objects around which the theory is transformed by a discrete $\IZ_n$ holonomy, e.g. the 4d $\IZ_n$ strings). The interesting question of constructing other possible defects, and the possible mathematical structures underlying their classification (in analogy with K-theory for open string tachyons) is left for future work. However, we cannot refrain from pointing out that in the setup from section  \ref{sec:hetz2}, the basic symmetries and their breaking are precisely as in type I theory with extra brane-antibrane pairs. In particular, this suggests a classification of topological brane charges in 10d heterotic in terms of a $KO$-theory (dovetailing heterotic/type I duality), in this case associated a pair of bundles $SO(32+n)\times SO(n)$ (namely, the gauge bundle and the normal bundle of the 10d slice in the supercritical $(10+n)$-dimensional spacetime), which annihilate via {\em closed} string tachyon condensation.

Focusing back on the construction of $\IZ_n$ defects, let us consider again the analysis of parity in the bosonic theory in section \ref{sec:parity}. This $\IZ_2$ symmetry is embedded into a $U(1)$ rotating the 2-plane $(x^{25},x^{26})$. The critical vacuum is recovered by a tachyon background $T\sim ({\rm Im}\, w)^2$ c.f. (\ref{tachyon-polar}), with $w\equiv x^{25}+ix^{26}=|w|e^{i\theta}$, whose zero cuts out the slice $\sin\theta=0$. In order to describe a $\IZ_2$ defect transverse to another 2-plane, e.g. $(x^{23},x^{24})$, we write $z\equiv x^{23}+ix^{24}=|z|e^{i\varphi}$ and consider a closed string tachyon background vanishing at the locus $\sin\theta'=0$, with $\theta'=\theta-\frac 12 \varphi$. The remaining spacetime is still critical, with a new angular coordinate in $(x^{23},x^{24})$ given by $\varphi'\sim\varphi+\frac 12 \theta$. A rotation $\alpha$ in $\varphi'$ (keeping $\theta'$ fixed at the tachyon minimum) is secretly a rotation  $\delta\phi=\alpha$, $\delta\theta=\alpha/2$; hence, a full $2\pi$ rotation results in a $\pi$ rotation in $\theta$, i.e. a $\IZ_2$ parity operation.
Other similar examples will be discussed in section \ref{sec:fluxbranes}.

\section{Discrete gauge symmetries as quenched translations}
\label{sec:translations}

The set of discrete symmetries amenable to the quenched rotation construction in the previous section is limited. In this section we present a far more universal embedding of discrete symmetries into continuous ones, which are realized as continuous translations in an extra (supercritical) $\IS^1$ dimension. The extra $\IS^1$ is subsequently eaten up by closed string tachyon condensation, which breaks the KK $U(1)$ symmetry down to the discrete subgroup. The basic strategy is to use periodic tachyon profiles, c.f. (\ref{before-limit}) to yield condensation processes which are well-defined on $\IS^1$. As mentioned at the end of section \ref{sec:super-bosonic}, we do not mind giving up exact solvability of the worldsheet CFT, and rely on the main lesson that condensation truncates dynamics to the vanishing locus of the 2d potential energy\footnote{This is analogous to the applications of open string tachyon condensation in annihilation processes, even if they are not exactly solvable BCFTs.}. 

\subsection{The mapping torus}
\label{sec:mapping-torus}

The basic ingredient in the construction is the mapping torus, whose construction we illustrate in fairly general terms. Although we apply it in the string theory setup, most of the construction can be carried out in the quantum field theory framework\footnote{Incidentally, the mapping torus (a.k.a. cylinder) is widely used in the study of global anomalies \cite{Witten:1982fp,Witten:1985xe,AlvarezGaume:1985ex}. It would be interesting to explore possible connections with our physical realization.}; the ultimate removal of the extra dimensions by tachyon condensation is however more genuinely stringy.

Consider an $N$-dimensional theory on a spacetime $\IX_N$, and let $\Theta$ be the generator of a discrete gauge symmetry $\IZ_n$. We consider extending the theory to $\IX_N\times {\bf I}$, where ${\bf I}$ is a one-dimensional interval\footnote{This is easily implemented in the supercritical bosonic and heterotic $HO^{+(n)}$ theories; for the supercritical type 0 orbifolds decaying to type II we must add another extra dimension, which can be kept non-compact; for the $HO^{+(n)\, /}$ theory, the orbifolding breaks the translational invariance and there is no actual continuous KK gauge symmetry, so we do not consider it here.} parametrized by a coordinate $0\leq y \leq 2\pi R$. We subsequently glue the theories at $y=0$ and $y=2\pi R$, but up to the action of $\Theta$. The final configuration is the theory on $\IX_N$ non-trivially fibered over $\IS^1$. The fibration is locally $\IX_N\times \IR$, but there is a non-trivial discrete holonomy implementing the action of $\Theta$. For example, if $\Theta$ is a discrete large isometry, the (purely geometric) glueing is
\beqa
(x, y=0) \simeq (\Theta(x),y=2\pi R)
\eeqa

Returning to the general situation, the extra dimension produces a KK $U(1)$ gauge boson from the $N$-dimensional viewpoint. The orbit of the associated translational vector field $\partial_y$ clearly contains the discrete $\IZ_n$ transformations, which are thus embedded as a discrete subgroup $\IZ_N\subset U(1)$.
States of the theory $\IX_N$ transforming with phase $e^{2\pi i \, p/n}$ under the discrete symmetry generator $\Theta$ extend as states with fractional momentum $p/n$ (mod $\IZ$) along $\IS^1$.  We note that the minimal $U(1)$ charge unit is $1/n$. 

In the supercritical string theory construction, the extra dimension is removed by a tachyon profile with periodicity $2\pi R$, which truncates the theory to the slice $Y=0$ (mod $2\pi R$). For instance, in the bosonic string theory, we sketchily write\footnote{Note that the radius $R$ can be kept arbitrary; for instance, the operator can be made marginal by turning on a lightlike dependence and adjusting the $\beta$ coefficient appropriately.}
\beqa
T\sim \mu^2 \, \big[\, 1-\cos\, \left({\textstyle\frac{Y}{R}} \right)\, \big]\, =\, 2\mu^2 \sin^2\left({\textstyle \frac{Y}{2R}}\right)
\label{periodic-tachyon}
\eeqa
and similarly in other supercritical string theories. Concretely, we use the  heterotic $HO^{+(n)}$ theory (since the $HO^{+(n)\,/}$ breaks translational invariance in the extra dimensions) and take $T\sim \sin (\frac{Y}{2R})$, as in the LHS of (\ref{tachyon-het-1}); for type II extended as supercritical type 0 orbifold, we take $T\sim \sin (\frac{Y}{2R}) X'$, obtained from (\ref{typeii-tach}) by renaming $X\to Y$ and taking a suitable $k'\to 0$ limit.

Since the periodic tachyon profile only excites components of integer KK momentum, it mimics a breaking of the $U(1)$ symmetry by fields of integer charge. Normalizing the minimal charge to $+1$, the breaking is implemented by fields of charge $n$. Hence, the continuous symmetry is broken to a discrete $\IZ_n$ symmetry  in the slice $y=0$, i.e. to the $\IZ_n$ of the original theory at $\IX_N$.

\subsection{Sum over disconnected theories}
\label{sec:superimposing}

There is an alternative orbifold description of the above construction (hence, particularly well-suited for string theory), as follows. We start with the theory extended to a trivial product $\IX_N\times (\IS^1)'$, with $(\IS^1)'$ a circle of length $2\pi n R$, parametrized by a coordinate $y'$. Subsequently, we mod out by the discrete action $\Theta$ on $\IX_N$, accompanied by a shift $y'\to y'+2\pi R$. The unit cell under this action is an $\IS^1$ of length $2\pi R$, along which the theory is twisted by the action of $\Theta$, as in the previous section. Similarly, we consider a tachyon profile with periodicity $2\pi R$.

We may regard the quotient theory on $\IS^1$ as the set of $\IZ_n$ invariant configurations\footnote{In orbifold language, this corresponds to restricting to the untwisted sector. Twisted states stretching between different zero loci of the tachyon will disappear in the process of tachyon condensation, so they can be ignored in the discussion.} in the parent theory on $\IX_N\times (\IS^1)'$. In this description, states with charge $p$ under $\Theta$ have integer KK momentum along $(\IS^1)'$, while the tachyon profile has KK momentum multiple of $n$. 

This viewpoint leads to an interesting interplay with the description of discrete gauge symmetries as a `sum over disconnected theories' \cite{Hellerman:2010fv}. Let us describe the latter, in the 4d avatar described in appendix A of that reference. Consider a 4d gauge theory with a non-perturbative sector restricted to instanton numbers multiple of $n$. The theory can be described as a sum over $n$ disconnected theories (with unconstrained instanton sector) with rotating $\theta$ angle,  $\theta_k=\theta+2\pi \frac kn$; schematically, an amplitude mediated by instanton number $p$ configuration reads 
\beqa
&& \sum_{p\in\IZ} \, \frac 1n\, \sum_{k=0}^{n-1}\, \langle {\rm out} |\, (...) \exp{[\, i (\theta+2\pi {\textstyle{\frac kn}})p\,]}\,|{\rm in}\rangle\, =\, \\
&& =\, \sum_{p\in\IZ} \,  \big[\,{\textstyle \frac 1n\,\sum_{k=0}^{n-1} \, \exp {(\, 2\pi i{\textstyle \frac  {kp}n}\,)}}\, \big] \langle {\rm out} |\, (...) \exp{(\, i \theta p\,)}\,|{\rm in}\rangle\,  =\, \sum_{p\in n\IZ} \, \langle {\rm out} |\, (...) e^{ i \theta p}\,|{\rm in}\rangle\quad  \nonumber
\eeqa
The projection operator in the second expression reflects the existence of a $\IZ_n$ discrete gauge symmetry, rotating the $\theta$ angle; pictorially, mapping the $k^{th}$ disconnected theory to the $(k+1)^{th}$. The construction admits a straightforward generalization to other field theories or string models.

This `sum over theories' prescription is reproduced by the tachyon condensation on the orbifold of the theory on $\IX_N\times (\IS^1)'$, as follows. There is a tachyon profile with $n$ zeroes, located at $y'=0, 2\pi R,\ldots, 2\pi R (n-1)$ in $(\IS^1)'$. Tachyon condensation produces $n$ disconnected copies of the theory on $\IX_N$ differing by the action of $\Theta^k$, $k=0,\ldots, n-1$. These copies correspond to the different `theories' which coexist in the superposition (which in this language, is nothing but restricting to orbifold invariant amplitudes).

\subsection{Topological $\IZ_n$ defects and quenched fluxbranes}
\label{sec:fluxbranes}

A basic property of theories with  discrete $\IZ_n$ gauge symmetries is the existence of $\IZ_n$ charged defects, real codimension-2 objects around which the theory is transformed by a discrete $\IZ_n$ holonomy. This non-trivial behaviour of the theory around the $\IS^1$ surrounding the $\IZ_n$ defect, is identical to the fibration over $\IS^1$ in the mapping torus in the previous sections. This may be regarded as an underlying reason for the universality of the mapping torus construction, which applies to fairly general discrete symmetries (as opposed to those in section \ref{sec:rotations}). In this section we use this relation to construct tachyon condensation profiles which produce the $\IZ_n$ charged defects of the theory, generalizing section \ref{sec:defects} to the more universal mapping torus setup.   

The construction is very reminiscent of the fluxbranes\footnote{We use the term `fluxbrane' in the original sense of extended solutions with non-trivial (compactly supported) magnetic fields in their transverse dimensions, rather than in the recent use as branes carrying worldvolume magnetic flux \cite{Hebecker:2011hk}. }  in \cite{Costa:2000nw,Saffin:2001ky,Gutperle:2001mb,Uranga:2001dx}, as we now explain. In compactifications on an $\IS^1$, parametrized by a coordinate $y$ with period $2\pi R$, a fluxbrane is a brane-like configuration with two transverse real coordinates constructed as follows. Pick a 2-plane, denoted by $(x^8, x^9)$ for concreteness, and use polar coordinates $z\equiv x^8+ix^9\equiv |z|e^{i\varphi}$. The fluxbrane configuration is obtained by performing dimensional reduction not along the original Killing vector $\partial_y$, but rather along $\partial_{y'}$, where $y'=y -\varphi R$ parametrizes a combined $\IS^1$. In the resulting solution, the orthogonal combination $\varphi'= \varphi+ y  R$ plays the role of angular coordinates in a 2-plane transverse to the fluxbrane, and there is a non-trivial magnetic flux  for the KK gauge boson $A_\mu\sim G_{\mu y'}$ (hence the name). In other words, circumventing the fluxbrane by rotating $\varphi ' $ secretly rotates in $\varphi$ and translates along $y$, such that the total holonomy is a whole shift in the original $\IS^1$ parametrized by $y$. 

The $\IZ_n$ defects can be constructed by using the same strategy in dimensional quenching, rather than in dimensional reduction. We start with the mapping torus of $\IX_N$ fibered over a $\IS^1$ parametrized by $y$, c.f. section \ref{sec:mapping-torus}. We choose a 2-plane $(x^8,x^9)$, or $z\equiv x^8+ix^9\equiv re^{i\varphi}$. Finally, we turn on a closed string tachyon profile (\ref{periodic-tachyon}), but now depending on $y'=y+\varphi R$, to remove one extra $\IS^1$ dimension. The resulting configuration contains a $\IZ_n$ defect at the origin of the 2-plane, since a rotation in $\varphi'$ results in a $\IZ_n$ holonomy. Note that the disappearance the KK gauge bosons in dimensional quenching (as compared with dimensional reduction, recall  section \ref{sec:rec-quench}), implies that there is no actual magnetic flux on the 2-plane, yet there is a non-trivial holonomy, as required to describe a $\IZ_n$ charged object. We refer to these $\IZ_n$ defects as `quenched fluxbranes'.
  
Note that in contrast with actual fluxbranes, we do not require quenched fluxbranes to solve the equations of motion of the spacetime effective theory. Instead, we use the construction to characterize the relevant topology describing $\IZ_n$ defects.

The construction makes manifest that $\IZ_n$ defects are conserved modulo $n$. Indeed, $n$ $\IZ_n$ defects correspond to a fluxbrane with trivial monodromy: going around it once implies moving $n$ times around $\IS^1$ in the mapping torus. The configuration can be trivialized by a coordinate reparametrization.
 
\subsection{Examples}
\label{sec:examples}

\subsubsection{Spacetime parity revisited}
\label{parity-revisited}

As an example, consider the realization of a spacetime $\IZ_2$ parity in e.g. the bosonic theory. Differently from section \ref{sec:parity}, the 27d supercritical geometry has the dimension $x^{25}$ fibered non-trivially along an extra $\IS^1$ parametrized by $y$, forming a M\"obius strip. In this non-orientable geometry, spacetime parity is a $\IZ_2$ subgroup of a continuous KK $U(1)$. The symmetry breaking is triggered by closed string tachyon  condensation. The $\IZ_2$ defects of the 26d theory, regions around which spacetime parity flips, can be constructed as quenched fluxbranes with a tachyon condensate (\ref{periodic-tachyon}), with the replacement $Y\to Y'=Y-R\varphi$, where $\varphi$ is the angle in the 2-plane transverse to the defect. 

\subsubsection{$\IZ_2$ symmetries of heterotic theories}
\label{heterotic-revisited}

We can easily implement the mapping torus construction to realize continuous versions of certain discrete symmetries of 10d heterotic theories. In order to have an extra (translational invariant) $\IS^1$, rather than an orbifold, we exploit the $HO^{+(1)}$ theory c.f. section \ref{sec:super-heterotic}. For instance, we can propose a different $U(1)$ embedding\footnote{Incidentally, the same discrete symmetry may have different supercritical embeddings into continuous symmetries. This is similar to embedding the same $\IZ_n$ symmetry into different continuous $U(1)$ groups, in fixed dimension.} of the $\IZ_2$ symmetry of the 10d $SO(32)$ theory of section \ref{sec:hetz2}, as follows. To reproduce the $\IZ_2$ holonomy along the $\IS^1$,  the supercritical  $HO^{+(1)}$ theory should have an $\IZ_2$ Wilson line introducing a $-1$ phase on $SO(32)$ spinors, e.g. 
\beqa
A={\textstyle\frac{1}{R}}\,\diag (\, i\sigma_2, 0,\ldots,0)\quad \to \quad \exp\Big({\textstyle\frac 12} \int_{\IS^1}  i\sigma_2 \, dy\Big)=-{\bf 1}
\eeqa
where the factor $\frac 12$ corresponds to the charge of spinors.

Consider a second example, given by the $\IZ_2$ symmetry exchanging the two $E_8$'s in the 10d $E_8\times E_8$ heterotic, i.e. an outer automorphism. The gauge nature of this symmetry, argued in \cite{Dine:1992ya}, can be made manifest using the mapping torus construction in the supercritical $E_8\times E_8$ theory mentioned in section \ref{sec:super-heterotic}. In this case, we must introduce a permutation Wilson line along the $\IS^1$, similar to those in CHL strings in lower dimensional compactifications \cite{Chaudhuri:1995fk,Chaudhuri:1995bf} (see also \cite{Ibanez:1987xa,Aldazabal:1995cf,Aldazabal:1994zk} for permutation Wilson lines in toroidal orbifolds).

\subsubsection{Discrete isometries of $\IT^2$}
\label{ex-tori}

We now give new examples, based on the discrete isometries of $\IT^2$ in appendix \ref{sec:tori-quintic}. To prevent a notational clash, we use $\beta\simeq \beta+2\pi$ to parametrize the supercritical $\IS^1$.

The construction of the mapping torus for $\IT^2$ is basically an orbifold of $\IT^3=\IT^2\times \IS^1$, by a rotation in $\IT^2$ and a simultaneous shift in $\IS^1$. They are described by free worldsheet CFTs and are familiar in orbifold constructions (in critical strings), see e.g. \cite{Dabholkar:2002sy}. Instead, we recast the construction in a language which will admit an easy generalization to CYs in projective spaces. In order to exploit the power of complex geometry, let us extend $\IS^1$ to $\IC^*\equiv \IC-\{0\}$, by introducing a variable $w=|w|e^{i\beta}$ (which can eventually be fixed to $|w|=1$ to retract onto $\IS^1$). The mapping torus is associated to an elliptic fibration over $\IC^*$, with constant $\tau$ parameter on the fiber, and suitable $SL(2,\IZ)$ monodromies around the origin. It turns out that holomorphic fibrations suffice for our purposes. Indeed, the constant $\tau$ holomorphic fibrations discussed in the context of F-theory \cite{Sen:1996vd,Dasgupta:1996ij}, can be readily adapt to the present (non-compact) setup.

Consider a Weierstrass fibration over a complex plane $w$
\beqa
y^2=x^3+f(w) x+g(w)
\eeqa
Our base space is non-compact, so we do not fix the degrees of the polynomials $f$, $g$.  From (\ref{jei}), a constant $\tau$ fibration is achieved by \cite{Sen:1996vd}
\beqa
f(w)=\alpha\, \phi(w)^2\quad ,\quad g(w) =\phi(w)^3
\eeqa
with $\phi(z)$ some polynomial. The value of $\tau$ is encoded in $\alpha$. 

In order to describe the $\IZ_2$ in table \ref{sec:tori-quintic}, which exists for generic values of $\tau$, we simply choose $\phi(w)=w$, and have
\beqa
y^2=x^3+\alpha \, w^2 \, x\,+\, w^3
\label{z2string}
\eeqa
Moving along $\IS^1$ (namely, $w\to e^{i\delta\beta}w$), the coordinates transform as $x\to e^{i \delta\beta } x$, $y\to e^{3i \delta\beta /2}y$. The  holonomy along $\IS^1$ is $x\to x$, $y\to -y$, precisely the desired $\IZ_2$ action. 

The construction of $\IZ_2$ defects is now straightforward. We simply introduce a complex coordinate $z$ for the two real transverse dimensions, and consider the configuration obtained 
from (\ref{z2string}) by the replacement $w\to w+z$. Note that two $\IZ_2$ strings are described by a fibration with $\phi=w^2$, which can be made trivial by a reparametrization, 
\beqa
y^2=x^3+\alpha \, w^4 \, x\,+\, w^6 \quad \longrightarrow\quad  y'^2=x'^3+\alpha x'+1\quad {\rm with }\ \ y=w^3 y ', \ x=w^2 x'
\label{annihilation}
\eeqa

\medskip

A similar discussion can be carried out for the remaining holomorphic $\IZ_n$ actions in appendix \ref{sec:tori-quintic}. Skipping further details, we simply quote the relevant fibrations:
\beqa
& \IZ_4 \; : \quad & y^2=x^3+w\,x\nonumber \\
& \IZ_6 \; : \quad & y^2=x^3+w \nonumber \\
& \IZ_3 \; : \quad & y^2=x^3+w^2 
\eeqa
These fibrations differ slightly from those in \cite{Dasgupta:1996ij}, because the latter describe crystallographic actions on global geometries. We note that the local behavior of their fibrations around fixed points on the base is equivalent to ours, modulo reparametrizations describing creation/annihilation of $n$ $\IZ_n$ defects.

\subsubsection{Discrete isometries in CYs: the quintic}

The above strategy generalizes easily to more general CYs, as we illustrate for the quintic ${\IX}_6={\IP}_5[5]$. Recall its expression (\ref{quintic}) at the Fermat point, 
\beqa
z_1^5+z_2^5+z_3^5+z_4^5+z_5^5=0
\eeqa
We simply focus on the $\IZ_5$ generated by $z_1\to e^{2\pi i/5}z_1$, with $z_2,\ldots, z_5$ invariant. The fibration associated to the mapping torus can be written
\beqa
w\, z_1^5+z_2^5+z_3^5+z_4^5+z_5^5=0
\label{quintic-string}
\eeqa
Setting $w=e^{i\delta y}$, moving along the $\IS^1$ gives $w\to e^{i\delta y} w$ and $z_1\to e^{-i \delta y/5} z_1$, so that completing the circle implements the desired $\IZ_5$ monodromy. The construction of $\IZ_5$ strings in the 4d theory amounts to a reinterpretation of $w$ in terms of the transverse coordinates, as in the previous section.

Note the important point that motion in $w$ does not correspond to changing the moduli of $\IX_6$. Recall that complex structure moduli are described by deformations of the defining equation corresponding to monomials  $\prod_i(z_i)^{n_i}$'s with $n_i<4$. This means that as one moves around the string there is no physical scalar which is shifting. This is fine because the monodromy is not part of a continuous gauge symmetry acting on any scalar of the 4d theory (as it disappears from the theory in the tachyon condensation).

These discrete isometries are relevant, since they often correspond to discrete R-symmetries of the 4d effective theory. The discussion of possible applications of our tools to phenomenologically interesting discrete R-symmetries is beyond our scope. 

\subsubsection{Antiholomorphic $\IZ_2$ and CP as a gauge symmetry}
\label{sec:cp}

A final class of discrete isometries of CY compactifications are given by antiholomorphic $\IZ_2$ actions, e.g. $z_i\to {\ov z_i}$, which are large isometries of the CY spaces with defining equations with real coefficients. These are orientation-reversing, and hence are not symmetries of the superstrings, but can be actual symmetries if  combined with an extra action. For instance, their combination with 4d parity gives a discrete symmetry, which in heterotic compactifications corresponds to a CP transformation \cite{Strominger:1985it}. Applying the mapping torus construction to this $\IZ_2$ symmetry (i.e. combining ingredients of the previous section and section \ref{parity-revisited}) results in a description of CP as a discrete gauge symmetry explicitly embedded in a (supercritical) $U(1)$ symmetry. This is a new twist in the history of realizing CP as a gauge symmetry, see e.g. \cite{Dine:1992ya,Choi:1992xp}.

\subsubsection{$\IZ_n$ symmetries already in $U(1)$ groups}
\label{sec:already-in-cont}

Although we have focused on discrete symmetries from (large) isometries, the constructions can be applied to general $\IZ_n$ discrete symmetries, even those embedded in continuous $U(1)$ factors already in the critical string theory (see \cite{BerasaluceGonzalez:2011wy,Ibanez:2012wg,BerasaluceGonzalez:2012vb,BerasaluceGonzalez:2012zn,Anastasopoulos:2012zu,Honecker:2013hda,fernando} for such symmetries in string setups). Focusing on the 4d setup for concreteness, recall \cite{Banks:2010zn} (also, the appendix in \cite{BerasaluceGonzalez:2012zn}) that the key ingredient is a $U(1)$ group with potential $A_1$, acting on a real periodic scalar $\phi\simeq \phi+1$ as
\beqa
A_1\to A_1+d	\lambda \quad , \quad \phi\to \phi +n\lambda
\eeqa
The $U(1)$ is broken, with $\phi$ turning into the longitudinal component of the massive gauge boson. But there is an unbroken $\IZ_n$, preserved even by non-perturbative effects. For instance, gauge invariance forces the amplitude of an instanton at a point $P$ to be dressed as
\beqa
e^{-2\pi i\phi}\, \exp(2\pi in\int_L A_1)
\label{inst}
\eeqa
which describes the emission of electrically charged particles of total charge $n$ (i.e. preserving the $\IZ_n$) along semi-infinite worldlines $L$ starting at $P$.

Let us now consider embedding this $\IZ_n$ symmetry as a mapping torus construction in a supercritical extension of the theory.  Along the extra $\IS^1$ there is a non-trivial $U(1)$ transformation (integrating to the $\IZ_n$ generator) and a corresponding shift $\phi\to \phi+1$. In other words there is one unit of flux for the field strength 1-form $F_1=d\phi$ 
\beqa
\int_{\IS^1} F_1=1
\label{flux1}
\eeqa
Notice that the mapping torus of length $2\pi R$ and the $n$-cover circle of length $2\pi nR$ (c.f. section \ref{sec:superimposing}) provide a physical realization of the two $\IS^1$'s in \cite{BerasaluceGonzalez:2012vb}, associated to the periodicity of $\phi$ and of the $U(1)$.

Fields with charge $q$ under the $\IZ_n$ have $\IS^1$ boundary conditions twisted by $e^{2\pi i\, q/n}$, hence have KK momenta $k+q/n$, with $k\in \IZ$, and so carry charge under the KK $U(1)$. This piece allows to recover (\ref{inst}) in this picture, as follows. The operator $e^{-2\pi i\phi}$ at a point in the $\IS^1$ (and at point $P$ in the critical spacetime) picks up phase rotations under translation, i.e. under a KK $U(1)$ transformation, which must be cancelled by those insertions of KK modes, with total KK momentum 1, e.g. $n$ states of minimal $\IZ_n$ charge.

The discussion in the previous paragraph has a nice string theory realization in the context of $\IZ_n$ symmetries arising from the $U(1)$ gauge groups on D-branes. In particular we focus on D6-brane models c.f. \cite{BerasaluceGonzalez:2011wy}, where $A_1$ is the gauge field on D6-branes, $\phi$ is the integral of the RR 3-form $C_3$ over some 3-cycle $\Sigma_3$, and the instanton is an euclidean D2-brane on  $\Sigma_3$. The non-trivial shift of $\phi$, namely the flux (\ref{flux1}), corresponds to a 4-form field strength flux
\beqa
\int_{\Sigma_3\times \IS^1} F_4=1
\eeqa
The D2-brane instanton on $\Sigma_3$ is not consistent by itself, but must emit particles with one unit of total KK momentum. This is more clear in the T-dual picture, which contains a D3-brane on $\Sigma_3\times \IS^1$ with one unit of $F_3$ flux over $\Sigma_3$, which must emit a fundamental string with one unit of winding charge \cite{Witten:1998xy}.

\section{Dual versions}
\label{sec:dual}

In this section we focus on yet another class of discrete large symmetries, which involve stringy non-geometric transformations. The corresponding mapping torus constructions can be regarded as containing non-geometric fluxes, in analogy with the duality twists in e.g. \cite{Dabholkar:2002sy}.

\subsection{Quantum symmetry in orbifolds}

Consider an orbifold compactification with parent space $\IY_6$ and orbifold group $\IZ_n$ with generator $\theta$ (which may also act on the gauge bundle in heterotic models). A general result from the structure of worldsheet amplitudes is the existence of a `quantum'\footnote{It is quantum in the sense of the worldsheet $\alpha'$ expansion. The symmetry is not visible in the large volume limit, because blow-up modes are in twisted sectors and thus $\IZ_n$-charged, so their vevs break the symmetry.} $\IZ_n$ discrete symmetry (different from the $\IZ_n$ generated by $\theta$). The generator $g$ acts on fields in the $\theta^k$-twisted sector with a phase $e^{2\pi i \, k/n}$. Invariance under this $\IZ_n$ leads to restrictions in amplitudes usually known as `point group selection rules', and have been discussed in the context of heterotic compactifications (see \cite{Bailin:1999nk,Ibanez:2012zz} for reviews, and e.g. \cite{Casas:1988vk,Kobayashi:2006wq} for early and recent applications). These symmetries have not been embedded in a continuous group.

For the present purposes, we may focus on the simple non-compact setup of $\IC^3/\IZ_n$, and take $\IC^3/\IZ_3$ as illustrative example. In fact, the quantum symmetry in this case is part of the $\Delta_{27}$ discrete symmetry analyzed in  \cite{Gukov:1998kn} (see \cite{Burrington:2006uu} for generalizations involving $\IZ_n$). It is straightforward to apply the mapping torus construction in section \ref{sec:translations} to derive this symmetry from a continuous $U(1)$ acting on an extra $\IS^1$. The construction is however particularly interesting because  it involves a non-geometric $\IZ_n$ action (see the next section for other examples).

It is worthwhile to point out that the quantum symmetry can be geometrized by application of mirror symmetry \cite{Hori:2000kt}. For instance, focusing on $\IC^3/\IZ_3$, the mirror geometry is described in e.g. \cite{Hanany:2001py} (in the setup of  D3-branes at singularities), and can be recast as
\beqa
&& z=uv \nonumber \\
&& y^2\, =\, x^3+zx+1
\eeqa
The geometry is a double fibration over the complex plane $z$, with one $\IC^*$ fiber parametrized by $u,v$ (degenerating over $z=0$), and the second fiber being a $\IT^2$ (c.f. appendix \ref{sec:tori}), degenerating over three points on the $z$-plane.

The mirror of the quantum $\IZ_3$ symmetry is generated by 
\beqa
x\to e^{2\pi i/3} x\quad ,\quad z\to e^{-2\pi i/3} z\quad ,\quad u\to e^{-2\pi i/3} u
\eeqa
This preserves the holomorphic 3-form $\Omega = \frac{dx}{y}\wedge \frac{du}{u}\wedge dz$. Note that the action restricted to $z=0$ is given by the $\IZ_3$ isometry of $\IT^2$ in appendix \ref{sec:tori}. 

Therefore, mirror symmetry renders the embedding of this $\IZ_3$ symmetry similar to the examples analyzed in Section \ref{sec:examples}.
 
\subsection{T-duality in type II $\IT^2$ compactifications}

T-duality is a fundamental duality, i.e. a discrete transformation implying the equivalence of different string models, or different points in the moduli space of a single string model. In systems mapped to themselves by a T-duality transformation, the latter should manifest as a discrete symmetry of the configuration. For instance, the 26d closed bosonic string theory on an $\IS^1$ of radius $R=\sqrt{\alpha'}$ is self-T-dual. In this case, the $\IZ_2$ symmetry is actually part of an enhanced $SU(2)^2$ gauge symmetry at the critical radius \cite{Giveon:1994fu,Polchinski:1998rq}.

For 10d type II theories we can obtain self-T-dual configurations by considering an square $\IT^2$ with equal radii $R=\sqrt{\alpha'}$, and vanishing B-field, i.e. $T=B+iJ=i$. This configuration is fixed by a $\IZ_4$ subgroup of the $SL(2,\IZ)$ duality group, generated by
\beqa
\begin{pmatrix} 0 & -1 \cr 1 & 0\end{pmatrix}
\label{t-dual}
\eeqa
However, there is no enhanced continuous gauge symmetry at this point, and the gauge nature of the $\IZ_4$ symmetry is not as manifest as in the bosonic case. 

This is however easily achieved in terms of a mapping torus configuration in a supercritical extension of the theory. The $\IZ_4$ symmetry is thereby embedded in a continuous KK $U(1)$, making its gauge nature manifest.

Notice that the above matrix (\ref{t-dual}) corresponds to the $\IZ_4$ isometry of $\IT^2$ in appendix \ref{sec:tori}. Rather than a coincidence, this is because both systems are related by application of one-dimensional mirror symmetry (T-duality along one dimension), which exchanges the $\IT^2$ K\"ahler and complex structure parameters $T\leftrightarrow \tau$. Therefore the mapping torus construction in this section is dual to those considered in section \ref{ex-tori}.

We conclude with a final remark regarding the mapping torus construction for the $\IZ_4$ self-T-duality. It provides an example of non-geometric fluxes in extra supercritical dimensions. Namely, as the theory moves along the extra $\IS^1$ it suffers a non-geometric transformation, which is not a diffeomorphism, but is a symmetry of string theory. This is a particular instance of the  non-geometric (but locally geometric) $Q$-fluxes in \cite{Shelton:2005cf}, which have been concretely considered in orbifold language in e.g. \cite{Dabholkar:2002sy}. Clearly the appearance of non-geometric fluxes along the supercritical $\IS^1$ will occur in the mapping torus construction of any discrete symmetry described by a non-geometric symmetry on the underlying compactification space. 

\section{Non-abelian discrete gauge symmetries}
\label{sec:nonabelian}

In this section we explore the generalization of the above ideas to the non-abelian case (see \cite{Alford:1989ch,Alford:1990mk,Alford:1990pt,Alford:1991vr,Alford:1992yx,Lee:1994qg} for early references on non-abelian discrete gauge symmetries, and \cite{BerasaluceGonzalez:2012vb,BerasaluceGonzalez:2012zn,fernando} (also \cite{Gukov:1998kn,Burrington:2006uu}) for recent string theory realizations). 

Consider an $N$-dimensional theory $\IX_N$ with a discrete gauge symmetry group $\Gamma$ (in general, not realized as a subgroup of a broken continuous symmetry). We would like to add extra dimensions in a supercritical extension of the theory, such that $\Gamma$ is embedded in a continuous non-abelian group $G$.

A possibility is that $G$ is an isometry group acting on the extra dimensional space $\IY$, as $g:y\to g(y)$. We may consider the trivial product $\IX_N\times \IY$ and quotient by the simultaneous action of the subgroup $\Gamma$ on the theory $\IX_N$ and the space $\IY$, namely $\gamma:(x,y)\to (\gamma(x),\gamma(y))$, for $\gamma\in \Gamma$. 
If the action of G on $\IY$ leaves no fixed points, the quotient generalizes the mapping torus construction in section \ref{sec:superimposing}, in the sense that non-trivial loops in the quotient define circles along which the theory $\IX_N$ is twisted by the action of an element of $\Gamma$. Subsequently turning on a non-trivial tachyon background invariant under $\Gamma$ (and hence with zeroes of the worldsheet potential related by $\Gamma$) would truncate the theory back to the critical theory $\IX_N$ with the desired symmetry breaking. 

A simple example of this construction is obtained by considering $\IY$ to be the group manifold $G$ itself, with action given by e.g. right multiplication; but any other $\IY$ on which $G$ acts transitively suffices. A more important and subtle point is that the action of $G$ on $\IY$ defined above does not descend in general to a globally well-defined action in the quotient. However, if $\Gamma$ is a normal subgroup\footnote{Recall that a group $N$ is a normal subgroup of $G$, written $N \triangleleft  G$, if $g^{-1}ng\in N$ for all $n\in N$, $g\in G$.} of $G$, then a globally well defined action in the quotient can be constructed (see appendix \ref{cosets}). Unfortunately, a normal discrete subgroup of a path-connected Lie group $G$ necessarily belongs to the center of $G$, and is therefore abelian. Hence, we cannot embed the non-abelian discrete symmetry in a continuous isometry/symmetry acting on the extra dimensions, broken by tachyon condensation when truncating to the critical theory.  

Configurations with local group actions, which are however not symmetries of the system have appeared in the context of discrete gauge symmetries in \cite{BerasaluceGonzalez:2012vb}. They describe a non-abelian continuous symmetry which is broken and has become massive by gauging  a set of scalars. This broken continuous symmetry may have an unbroken discrete subgroup, which manifests as a discrete gauge symmetry of the theory. This perspective is useful to deal with the non-abelian discrete symmetry $\Gamma$ of the theory $\IX_N$, and its embedding into a continuous group $G$ acting on the quotient $(\IX_N\times \IY)/\Gamma$. This construction provides an embedding of $\Gamma$ into a continuous symmetry which is broken and made massive by a process of gauging. The tachyon condensation would then truncate the gauged theory to the critical spacetime slice, triggering no additional symmetry breaking. Note that the construction in section \ref{sec:already-in-cont} can be regarded as a particular realization of this idea in the abelian case.

\subsection{Discrete Heisenberg group from supercritical magnetized tori}
\label{heisenberg}

It is easy to provide concrete examples of this realization; in this section we present an embedding of a discrete Heisenberg group in terms of a supercritical extension with magnetized $\IT^2$, in a setup close to \cite{BerasaluceGonzalez:2012vb}. A more formal discussion in terms of cosets is left for appendix \ref{cosets}.

Consider the theory $\IX_N$ to have a $U(1)$ gauge symmetry and a discrete gauge symmetry, generated by two order-$n$ elements $A$, $B$ (with $A^n=B^n=1$) which commute to an element of $U(1)$, that is $AB=CBA$ with $C\in U(1)$ (note that $C$ is required to be of order $n$ as well). This is a discrete Heisenberg group, which we denote by $H_n$. We assume that $H_n$ is not embedded into any continuous (massive or not) symmetry of $\IX_N$.

Consider now a supercritical extension of the theory with two extra dimensions (two plus two for type 0 extensions of type II models), which parametrize a $\IT^2$ (for simplicity taken square with unit length coordinates $x,y$). In analogy with the mapping torus construction, we specify that the theory $\IX_N$ picks up the action of the generators $A$, $B$, as one moves along the two fundamental cycles of $\IT^2$. This embeds the two discrete generators into continuous translational $U(1)$ actions, the would-be KK gauge symmetries of the theory. These symmetries are however broken, even before the process of tachyon condensation. To see this, note that moving around the whole $\IT^2$ results in an action $ABA^{-1}B^{-1}=C$, namely there is a circulation of the $U(1)$ gauge potential. This implies that there is a non-trivial $U(1)$ magnetic field (with $n$ units of flux \cite{BerasaluceGonzalez:2012vb}) on the $\IT^2$. Although the field strength can be taken constant and the configuration seems translational invariant, the gauge potential can be written
\beqa
A\,=\, \pi n (xdy-ydx) 
\eeqa
so that translations in $x$ imply a change in the Wilson line of $A$ along $y$, and viceversa. This can be regarded as an action of the KK $U(1)$'s on the Wilson line scalars, which define a gauging that breaks the continuous symmetry and makes the KK gauge bosons massive \cite{BerasaluceGonzalez:2012vb,fernando}. There is however an unbroken $H_n\rtimes U(1)$ symmetry, corresponding to the original group of $\IX_N$.

We may subsequently turn on a non-trivial periodic tachyon profile to remove the extra dimensions, and truncate the dynamics to the origin. For instance, we may take the supercritical heterotic theory $HO^{+(2)}$ with $T^{33}\sim \sin (\pi x)$, $T^{34} \sim\sin (\pi y)$. We  emphasize that the tachyon background does not lead to any additional breaking of the symmetry. It should be straightforward to find other examples of non-abelian discrete symmetries embedded in continuous massive symmetries.

\section{Conclusions}
\label{sec:final}

In this paper we have shown that genuinely discrete gauge symmetries in string theory can actually be embedded into continuous symmetries, by extending the theories beyond the critical dimension and using closed string tachyon condensation. We have discussed several different examples of such symmetries and embeddings, and their relationship. The examples include outer automorphisms of the gauge group, discrete isometries of the compactification space, quantum symmetries at orbifolds, etc. 

The construction brings these symmetries into a framework close to other discrete symmetries obtained as remnants of continuous ones, e.g. \cite{Banks:2010zn}; however the symmetry breaking mechanism by closed string tachyon condensation implies important novelties compared with the familiar gauging/Higgsing ones. Interestingly, the realization of discrete symmetries as quenched translations makes contact with the alternative description of discrete symmetries as a sum over disconnected theories in \cite{Hellerman:2010fv}.

We have also discussed aspects of the generalization to non-abelian discrete symmetries, which typically require invoking an additional gauging mechanism. It would be interesting to exploit our constructions to study this and more general cases, and gain insights towards a general picture of discrete symmetries in string theory.

The most relevant gain achieved by embedding the discrete symmetries into continuous ones is the realization of charged topological defects as tachyon solitons. This is very reminiscent of the realization of D-branes as open string tachyon solitons, and K-theory. An important novelty is that our construction, for instance for large isometries, can produce objects with charges associated to the gravitational (rather than RR) sector of the theory. Hopefully, future work will clarify the mathematical structures underlying closed string tachyon condensation and its topological solitons. We hope to return to this and other related interesting points in the future.

\bigskip

\centerline{\bf \large Acknowledgments}

\bigskip

We thank P. G. C\'amara, A. Collinucci, L. Ib\'a\~nez, I. Garc\'ia-Etxebarria, I. Valenzuela, J.M. Nieto and especially F. Marchesano and D. Regalado for discussions.  
This work has been supported by the Spanish Ministry of Economy and Competitiveness under grants FPA2009-09017, FPA2009-07908, FPA2010-20807, FPA2012-32828, Consolider-CPAN (CSD2007-00042),  the grants  SEV-2012-0249 of the Centro de Excelencia Severo Ochoa Programme,   HEPHACOS-S2009/ESP1473 from the C.A. de Madrid, SPLE Advanced Grant under contract ERC-2012-ADG$\_$20120216-320421 and the contract ``UNILHC" PITN-GA-2009-237920 of the European Commission. M.B-G. acknowledges the finantial support of the FPU grant AP2009-0327. M.M acknowledges support from the Excellence Campus M. Sc. in Theoretical Physics scholarship.

\newpage

\appendix

\section{Partition functions of supercritical strings}
\label{sec:partition-functions}

Despite the long history of supercritical strings \cite{Chamseddine:1991qu}, their detailed construction may be unfamiliar to most readers. In this appendix we complement the main text references therein (e.g. \cite{Hellerman:2004zm,Hellerman:2006nx}) by providing their partition functions, which help in visualizing the diverse GSO-like projections.
Their construction is standard, with the linear dilaton simply readjusting the zero point energy to reproduce the masses used in the main text.

We start with the supercritical bosonic string in $D$ dimensions, for which we have
\beqa
Z(\tau)\,=\, \frac{(4\pi^2\alpha'\tau_2)^{-\frac{D-2}{2}}}{\, |\eta|^{2(D-2)} }
\eeqa
For the supercritical type 0A/B in $D=10+n$ dimensions, the partition function reads

\beqa
Z(\tau)&\!\!=&\!\!\frac{1}{2}\frac{(4\pi^2\alpha'\tau_2)^{-4-\frac{n}{2}}}{|\eta|^{16+2n}}\frac{1}{|\eta|^{8+n}}
\bigg( \Big|\vartheta{\scriptsize \Big[ \begin{array}{c} \!\! 0 \!\! \\ \!\! 0\!\! \end{array} \Big]}\Big|^{8+n}\!\!+ \Big|\vartheta{\scriptsize \Big[ \begin{array}{c} \!\! 0 \!\! \\ \!\! 1/2\!\! \end{array} \Big]}\Big|^{8+n}\!\! + \Big|\vartheta{\scriptsize \Big[ \begin{array}{c} \!\! 1/2 \!\! \\ \!\! 0 \!\! \end{array} \Big]}\Big|^{8+n}\!\! \mp\Big|\vartheta{\scriptsize \Big[ \begin{array}{c} \!\! 1/2 \!\! \\ \!\! 1/2 \!\! \end{array} \Big]}\Big|^{8+n} \bigg)\nonumber \\
&=& \frac{1}{2}\frac{(4\pi^2\alpha'\tau_2)^{-4-\frac{n}{2}}}{|\eta|^{16+2n}} \left( \left|  Z^{0 }_{0 } \right|^{8+n} + \left|  Z^{1 }_{0 } \right|^{ 8+n }  + \left|  Z^{ 0}_{1 } \right|^{8+n} \mp \left|  Z^{1 }_{1 } \right|^{ 8+n}  \right)
\eeqa
In the last line we have introduced the notation from \cite{Polchinski:1998rr}
\begin{equation}
 Z^{a}_{ b} (\tau )  =  \dfrac{\vartheta { a/2\brack b/2} }{ \eta (\tau ) }
\end{equation} 
Consider now type 0 supercritical theories in $D=10+2k$ dimensions, and orbifold by the $\IZ_2$ flipping $k$ extra dimensions, and worldsheet fermions as described in the text. The partition function sums over untwisted and twisted sectors, and reads 
\beqa
&&\!\!\!\!\!\!\!\!Z_{0}^{\rm orb}(\tau  ) =\frac{1}{4} (4\pi^2 \alpha ' \tau_2 )^{-4} \mid \eta  \mid^{-16} \times \nonumber \\
&&\times   \left[ (4\pi^2 \alpha ' \tau_2 )^{-n/2} \mid \eta  \mid^{-2n} \left( \left|  Z^{0 }_{0 } \right|^{8  } \left|  Z^{0 }_{0 } \right|^{n  } + \left|  Z^{1 }_{0 } \right|^{ 8 } \left|  Z^{1 }_{0 } \right|^{ n } + \left|  Z^{ 0}_{1 } \right|^{8  } \left|  Z^{0 }_{1 } \right|^{n  } \mp \left|  Z^{1 }_{1 } \right|^{ 8 } \left|  Z^{1 }_{1 } \right|^{n  } \right) \right.\nonumber \\
&&\quad \left. +  \mid Z^{1}_0  \mid^{-n} \left( \left|  Z^{0 }_{0 } \right|^{8  } \left|  Z^{0 }_{1 } \right|^{n  } + \left|  Z^{1 }_{0 } \right|^{ 8 } \left|  Z^{1 }_{1 } \right|^{ n } + \left|  Z^{ 0}_{1 } \right|^{8  } \left|  Z^{0 }_{0 } \right|^{n  } \mp \left|  Z^{1 }_{1 } \right|^{ 8 } \left|  Z^{1 }_{0 } \right|^{n  } \right) \right. \nonumber \\
&&\quad \left. + \mid Z^{0}_1   \mid^{-n} \left( \left|  Z^{0 }_{0 } \right|^{8  } \left|  Z^{1 }_{0 } \right|^{n  } + \left|  Z^{1 }_{0 } \right|^{ 8 } \left|  Z^{0 }_{0 } \right|^{ n } + \left|  Z^{ 0}_{1 } \right|^{8  } \left|  Z^{1 }_{1 } \right|^{n  } \mp \left|  Z^{1 }_{1 } \right|^{ 8 } \left|  Z^{0 }_{1 } \right|^{n  } \right) \right.\nonumber  \\
&&\quad \left. + \mid Z^{0}_0   \mid^{-n} \left( \left|  Z^{0 }_{0 } \right|^{8  } \left|  Z^{1 }_{1 } \right|^{n  } + \left|  Z^{1 }_{0 } \right|^{ 8 } \left|  Z^{0 }_{1} \right|^{ n } + \left|  Z^{ 0}_{1 } \right|^{8  } \left|  Z^{1 }_{0 } \right|^{n  } \mp \left|  Z^{1 }_{1 } \right|^{ 8 } \left|  Z^{0 }_{0 } \right|^{n  } \right) \right]
\eeqa
This can be  written compactly as
\begin{align}Z^{orb.}_0=\frac{(4\pi^2\alpha'\tau_2)^{-4}}{\vert\eta(\tau)\vert^{16}}\dfrac{1}{4}\sum_{a,b,c,d=0}^1Z_{abcd}.\end{align}
by defining generalized `Z' functions for different sectors 
\beqa
Z_{ab;cd}=|\eta|^{-8}\Bigg|\vartheta{\footnotesize \bigg[ \begin{array}{c} \!\! \frac 14(1+(-)^{a}) \!\! \\ \!\! \frac 14 (1+(-)^{c})\!\! \end{array} \bigg]}\Bigg|^{8}\;
\Bigg|\vartheta{\footnotesize \bigg[ \begin{array}{c} \!\! \frac 14(1+(-)^b) \!\! \\ \!\! \frac 14 (1+(-)^d)\!\! \end{array} \bigg]_B}\Bigg|^{-n}\;
\Bigg|\vartheta{\footnotesize \bigg[ \begin{array}{c} \!\! \frac 14(1+(-)^{a+b}) \!\! \\ \!\! \frac 14 (1+(-)^{c+d})\!\! \end{array} \bigg]}\Bigg|^{n}\;
\label{generalized}
 \eeqa
Note that each bosonic thetas (indicated with`$B$') must be replaced by  $\eta(\tau)^{3}(4\pi^2\alpha'\tau_2)^{-1/2}$ in the $b=d=0$ sector.

Let us move on to the heterotic theories. For the $HO^{(n)}$ theory, the partition function is
\begin{equation}
\begin{split}
\! \! \! \! \! \! \! \! \! \! \! \! \! \! \! \! \! \! \! \! \! \! Z_{HO^{+(n)}}(\tau  ) =\dfrac{1}{4}\frac{ (4\pi^2 \alpha ' \tau_2 )^{-4-\frac{n}{2}} }{\mid \eta  \mid^{16+2n}}   \times \left[ \left(  Z^{0 }_{0 } \right)^{16  } + \left(  Z^{1 }_{0 } \right)^{16  } + \left(  Z^{0 }_{1 } \right)^{16  } + \left(  Z^{1 }_{1 } \right)^{16  } \right] \\
 \times \left[ \left(  \bar{Z}^{0 }_{0 } \right)^{4  } \left|  Z^{0 }_{0 } \right|^{ n  } - \left( \bar{Z}^{1 }_{0 } \right)^{4  }  \left|  Z^{1 }_{0 } \right|^{ n  }   - \left( \bar{Z}^{0 }_{1 } \right)^{4  }  \left|  Z^{0 }_{1 } \right|^{ n  } \pm \left( \bar{Z}^{1 }_{1 } \right)^{4  }  \left|  Z^{1 }_{1 } \right|^{ n  } \right] 
\end{split}
\end{equation}
Note the familiar $SO(32)$ sector in the last factor of the first line. The familiar change of GSO projections in these 32 fermions can be used to directly construct the $E_8\times E_8$ version of the theory used in section \ref{heterotic-revisited}.

For the $HO^{+(n)\, /}$ theory, the partition function is
\begin{align}
 &Z_{HO^{+(n)/}}(\tau)=\dfrac{1}{2} \frac{ (4\pi^2 \alpha ' \tau_2 )^{-4-\frac{n}{2}} }{\mid \eta  \mid^{16+2n}} \times \\
&\times\!\bigg[\! \left(\bar{Z}^{0}_{0} \right)^{4}\left(Z^{0}_{0}\right)^{16}\left|Z^{0}_{0}\right|^{n}\! -\left( \bar{Z}^{1}_{0}\right)^{4}\left(Z^{1}_{0}\right)^{16}\left|Z^{1}_{0}\right|^{n} \! -\left(\bar{Z}^{0}_{1}\right)^{4} \left(Z^{0}_{1}\right)^{16}\left|Z^{0}_{1}\right|^{n}\! \pm\left(\bar{Z}^{1}_{1} \right)^{4}\left(Z^{1}_{1}\right)^{16}\left|Z^{1}_{1}\right|^{n}\!\bigg]\nonumber 
\end{align}
The orbifold can be written compactly using the generalized functions (\ref{generalized}), as
\begin{align}
Z^{orb}_{HO^{(+n)/}}=\frac{(4\pi^2\alpha'\tau_2)^4}{\vert\eta(\tau)\vert^{16}}\frac{1}{4}\sum_{a,b,c,d=0}^1(-1)^{a+c}Z_{abcd}.
\end{align}

\newpage

\section{Some examples of discrete isometries}
\label{sec:tori-quintic}

In this Appendix we gather a few standard yet illustrative examples of discrete isometries corresponding to large diffeomorphisms  in compact manifolds. We first review the case $\IT^2$, and move on to the quintic. Extension to other CY hypersurfaces is straightforward.

\subsection{Discrete isometries of $\IT^2$}
\label{sec:tori}

The 2-torus can be described as a quotient $\IR^2/\Gamma$ of the 2-plane by a lattice $\Gamma$ of translations. Beyond the $U(1)^2$ continuous isometry group, there are possible discrete isometries from crystallographic symmetries of $\Gamma$, which correspond to large diffeomorphisms.  Introducing a complex coordinate $z$ with periodicities $z\simeq z+1$, $z\simeq z+\tau$, these isometries are subgroups of the $SL(2,\IZ)$ modular group, leaving the lattice invariant  (possibly for some specific choice of $\tau$). An alternative description of $\IT^2$ is via the Weierstrass equation
\beqa
y^2=x^3+fx+g
\label{weierstrass}
\eeqa
The coordinates $x,y$ relate to $z$ by the so-called `uniformization mapping' (see e.g. \cite{dlmf}). Here $f,g$ are complex constants, in terms of which the $j$-function of the complex structure parameter $\tau$ is 
\beqa
j(\tau)=\frac{2(24 \,f)^3}{27g^2+4f^3}
\label{jei}
\eeqa
The discrete symmetries, and the values of the complex structure modulus at which they hold, are familiar from the construction of toroidal orbifolds. They are
{\small
\begin{center}
\begin{tabular}{|c|c|c|c|c|c|c|}
\hline
Symmetry &  $\tau$ & Generator &  $f$, $g$ & Weierstrass & Generators & $SL(2,\IZ)$  \\
\hline\hline
$\IZ_2$ & arbitr. & $z\to -z$ & arbitr. & $y^2=x^3+fx+g$ & \begin{tabular}{c} $x\to x,$ \cr $y\to -y$ \end{tabular} & $\begin{pmatrix} -1 & 0 \cr 0 & -1\end{pmatrix}$\\
\hline
$\IZ_4$ & $\tau=i$ & $z\to e^{2\pi i/4}z$ & $g=0$ & $y^2=x^3-x$ & \begin{tabular}{c} $x\to -x,$\cr $y\to iy$\end{tabular} & $\begin{pmatrix} 0 & -1 \cr 1 & 0\end{pmatrix}$ \\
\hline
$\IZ_6$ & $\tau=e^{\pi i/3}$ & $z\to e^{2\pi i/6} z$ & $f=0$ & $y^2=x^3+1$ & \begin{tabular}{c} $x\to e^{2\pi i/3} x,$ \cr $y\to -y$ \end{tabular} & $\begin{pmatrix} 1 & 1 \cr -1 & 0\end{pmatrix}$\\
\hline\hline
$\IZ_2$ & $\tau_1=1$ & $z\to {\ov z}$ &  real  &  $y^2=x^3+fx+g$ & \begin{tabular}{c} $x\to {\ov x}$\cr $y\to {\ov y}$ \end{tabular} &  $\begin{pmatrix} 1 & 0 \cr 0 & -1\end{pmatrix}$ \\
\hline
$\IZ_2$ & $\tau_1=\frac 12$ & $z\to {\ov z}$ &  real  &  $y^2=x^3+fx+g$ & \begin{tabular}{c} $x\to {\ov x}$\cr $y\to {\ov y}$ \end{tabular} &  $\begin{pmatrix} 1 & 0 \cr 1 & -1\end{pmatrix}$ \\
\hline
\end{tabular}
\end{center}
}
Here we have applied certain rescalings to simplify  the Weierstrass equation for $\IZ_4$ and $\IZ_6$. Note that, by squaring the  $\IZ_6$ generator, there is a $\IZ_3$ symmetry for $\tau=e^{\pi i/3}$. Finally, the last two entries correspond to orientation-reversing actions, and they are actually not in $SL(2,\IZ)$.

\subsection{Discrete isometries of the quintic}

Consider the quintic CY $\IX_6=\IP_5[5]$ at the Fermat point, i.e. the hypersurface with defining equation
\beqa
z_1^5+z_2^5+z_3^5+z_4^5+z_5^5=0
\label{quintic}
\eeqa
This has a discrete symmetry group  $(\IZ_5)^4 \times S_5$. The $S_5$ is the group of permutations of 5 elements, the  homogeneous coordinates $z_i$. The $\IZ_5$'s are generated by independent phase rotations  $z_i\to e^{2\pi i /5}z_i$, with the removal of an overall phase rotation which is part of the projective action to define the ambient $\IP_5$.

In addition, there is an antiholomorphic action 
\beqa
z_i \to {\ov z}_i
\eeqa
These actions, and their products, have been extensively exploited in the literature.

\newpage

\section{Non-abelian discrete symmetries and cosets}
\label{cosets}

We describe in more formal terms some of the ingredients in the embedding of non-abelian discrete symmetries into continuous groups. We start with a general framework description with broad applicability, which we subsequently apply to the discrete Heisenberg group in section \ref{heisenberg}.

\subsection{General framework}

Consider a theory $\mathcal{M}$, with a (right-acting) discrete symmetry group $\Gamma$, which we want to embed as part of a supercritical theory. We also have a group $G$ with a normal subgroup $N$. As a first step in the construction, we build the trivial fiber bundle $E\equiv G\times \mathcal{M}$, with canonical projection map onto the first factor $\pi(g,m)=g$. There is a canonical action $A$  of $G$, so that for each $g'\in G$, $A(g'):(g,m)\to (g'g,m)$.

We now pick a homomorphism $\phi: N\rightarrow \Gamma$ and for each $n\in N$ consider the action $F_n:(g,m)\to (ng, m\cdot\phi(g^{-1} ng))$. We quotient $E=G\times \mathcal{M}$ by the equivalence relation $e_1\sim e_2$ if $F_n(e_1)=e_2$ for some $n\in N$. The quotient, denoted by $E/F$, has a natural projection $\pi'$  onto $G/N$ with preimage $\mathcal{M}$, so the configuration is a non-trivial fiber bundle over $G/N$. Furthermore, thanks to the normality of $N$, the action $A(g)$ descends to a well-defined action in the quotient (namely, the images under $A(g)$ of $F$-equivalent points are $F$-equivalent). 

The topology of the bundle is specified by the holonomies around non-trivial loops in $G/N$. Consider a one-parameter curve $g(t)$ in $G$, with $t\in[0,1]$, going from the identity to some $n\in N$, namely $g(0)=1$, $g(1)=n$. This descends to a closed loop in $G/N$, with a non-trivial action on the fiber ${\cal M}$. In particular, a point of $E/F$ with representative $(g,m)\in E$ comes back as the point $(ng,m)\in E$, which is in the class of $(g,m\cdot\phi(g^{-1}n^{-1}g))$ in $E/F$. Namely, the fiber suffers a monodromy given by an element in $\phi(N)= \Gamma$, the discrete symmetry group.  The construction succeeds in embedding this symmetry as part of the continuous group $G$ acting on the coset $G/N$. Notice however, that the continuous group may act as a non-trivial shift of scalars, resulting in a gauging which makes the symmetry massive.

The last step would be to introduce a tachyon restricting the dynamics to the identity class in $G/N$. This is difficult to describe in general, but can be worked out in detail in examples.

\subsection{The Heisenberg group}

We now describe a particular example, which eventually corresponds to the Heisenberg group in section \ref{heisenberg}. This illustrates some of the ingredients ultimately leading to massive continuous symmetries.

We take $G$ to be the Heisenberg group $H_3(\IR)$ and introduce the normal subgroup $N$, which has  a normal subgroup $N_N$, i.e. $N_N\triangleleft N\triangleleft G$. They are defined by the sets of matrices of the form
\begin{align}
G=\left(\begin{array}{ccc}1&x&z+\frac{xy}{2}\\0&1&y\\0&0&1\end{array}\right),\;\; N=\left(\begin{array}{ccc}1&n_x&h_z+\frac{n_xn_y}{2}\\0&1&n_y\\0&0&1\end{array}\right),\;\;N_N=\left(\begin{array}{ccc}1&n_x&n_z+\frac{n_x n_y}{2}\\0&1&n_y\\0&0&1\end{array}\right), \nonumber
\end{align}
with $x,y,z,h_z\in\mathbb{R}$, $n_x,n_y,n_z\in\mathbb{Z}$. The quotients are $G/N\simeq \IT^2$, $N/N_N\simeq U(1)$. 

We consider a theory  $\mathcal{M}$ with a symmetry $H_n\rtimes U(1)$, with $H_n$ a discrete Heisenberg group. The action of $N$ on this theory (the homomorphism $\phi$ above) includes a $U(1)$ gauge transformation with parameter $e^{2\pi i h_z}$. Incidentally, note that the homomorphism $\phi$ has a non-trivial kernel, given by $N_N$. We now take the product $G\times \mathcal{M}$, and quotient by the equivalence relation $(g,e^{2\pi i\theta})\sim (hg,m\!\,\cdot\!\, \phi(g^{-1}h g), e^{2\pi i(h_z+\theta)})$, where the last entry describes the gauge $U(1)$ fiber, and $m$ represents other sectors of the $\mathcal{M}$ on which $\phi$ acts. The resulting quotient space is a non-trivial fiber bundle over $\IT^2$. The non-trivial glueing is manifest in the identifications $(x+1,y,z+\frac{y}{2})$, $(x,y+1,z-\frac{x}{2})$, so the $U(1)$ fiber transforms as $e^{2\pi i z}\to e^{2\pi i z} e^{i\pi y}$ under $x\to x+1$, and as $e^{2\pi i z}\to e^{2\pi i z} e^{-i\pi x}$ under $y\to y+1$.
A connection on this bundle must satisfy $A(x+1,y)=A(x,y)-\pi dy$, $A(x,y+1)=A(x,y)+\pi dx$, which imply that the solutions carry a nonzero magnetic flux.

The action of $G$ does not correspond to a true symmetry of the background, as follows. On $G\times \mathcal{M}$, it takes $(g, e^{2\pi i\theta})$ to $(g'g, e^{2\pi i\theta})$. To find the action on the quotient, we take without loss of generality $g=(x,y,0)$ and $g'=(a,b,c)$, and have 
\begin{align}
(g'g, e^{2\pi i\theta})\,=\,\big(x+a,y+b,c + \frac{1}{2}(ay-bx) , e^{2\pi i\theta}e^{\pi i(-2c+bx-ay+( a+x) y-( b+y) x)}\big).
\end{align}
It maps fibers at different points, and also acts by moving along the fiber, i.e. translations plus gauge transformations, as usual in the magnetized torus. The true symmetries are given by the  transformations $A(n)$ for $n\in N$, which are precisely $H_n\rtimes U(1)$, recovering the results in \cite{BerasaluceGonzalez:2012vb,fernando}.

\end{document}